\documentclass[aps,twocolumn,superscriptaddress,nofootinbib,a4paper,longbibliography]{revtex4-2}
\pdfoutput=1

\usepackage{times}
\usepackage{epsfig}
\usepackage{amsfonts}
\usepackage{amsmath}
\usepackage{amsthm}
\usepackage{amssymb}					
\usepackage{amsthm}
\usepackage{dsfont}
\usepackage{bm}
\usepackage{mathtools}
\usepackage{color}
\usepackage{multirow}
\usepackage[normalem]{ulem}
\newcommand{\stkout}[1]{\ifmmode\text{\sout{\ensuremath{#1}}}\else\sout{#1}\fi}
\usepackage{latexsym}
\usepackage{mathrsfs}
\usepackage{natbib}
\usepackage{verbatim}
\usepackage[T1]{fontenc}
\usepackage{float}
\usepackage{graphicx}
\usepackage{xcolor}
\usepackage{physics}
\usepackage{soul}

\theoremstyle{definition}

\newcommand{\bracket}[3]{\langle#1|#2|#3\rangle}
\newcommand{\expect}[1]{\langle#1\rangle}

\usepackage[colorlinks=true,linkcolor=blue,citecolor=magenta,urlcolor=blue]{hyperref}

\begin{document}

\title{Genuine multipartite entanglement detection with imperfect measurements: concept and experiment}

\author{Huan Cao}
\email{huan.cao@univie.ac.at}
\thanks{These two authors contributed equally to this work}
\affiliation{University of Vienna, Faculty of Physics, Vienna Center for Quantum Science and Technology (VCQ), 1090 Vienna, Austria}
\affiliation{Christian Doppler Laboratory for Photonic Quantum Computer, Faculty of Physics, University of Vienna, 1090 Vienna, Austria}

\author{Simon Morelli}
\email{smorelli@bcamath.org}
\thanks{These two authors contributed equally to this work}
\affiliation{BCAM - Basque Center for Applied Mathematics,
Mazarredo 14, E48009 Bilbao, Basque Country - Spain}

\author{Lee A. Rozema}
\affiliation{University of Vienna, Faculty of Physics, Vienna Center for Quantum Science and Technology (VCQ), 1090 Vienna, Austria}

\author{Chao Zhang}
\email{drzhang.chao@ustc.edu.cn}
\affiliation{CAS Key Laboratory of Quantum Information, University of Science and Technology of China, 230026 Hefei, China.}
\affiliation{CAS Center For Excellence in Quantum Information and Quantum Physics, University of Science and Technology of China, 230026 Hefei, China.}
\affiliation{Hefei National Laboratory, University of Science and Technology of China, 230088 Hefei, China}

\author{Armin Tavakoli}
\email{armin.tavakoli@teorfys.lu.se}
\affiliation{Physics Department, Lund University, Box 118, 22100 Lund, Sweden.}

\author{Philip Walther}
\affiliation{University of Vienna, Faculty of Physics, Vienna Center for Quantum Science and Technology (VCQ), 1090 Vienna, Austria}
\affiliation{Christian Doppler Laboratory for Photonic Quantum Computer, Faculty of Physics, University of Vienna, 1090 Vienna, Austria}

\date{\today}

\begin{abstract}
Standard procedures for entanglement detection assume that experimenters can exactly implement specific quantum measurements. Here, we depart from such idealizations and investigate, in both theory and experiment, the detection of genuine multipartite entanglement when measurements are subject to small imperfections. For arbitrary qubits number $n$, we construct multipartite entanglement witnesses where the detrimental influence of the imperfection is independent of $n$.  In a tabletop four-partite photonic experiment we demonstrate first how a small amount of alignment error can undermine the conclusions drawn from standard entanglement witnesses, and then perform the correction analysis.   Furthermore, since we consider quantum devices that are trusted but not perfectly controlled, we showcase advantages in terms of noise resilience as compared to device-independent models. 
\end{abstract}

\maketitle


 Verifying entangled states is a central challenge in quantum information~\cite{Horodecki2009,Friis2019}. Commonly, this is achieved by measuring a suitable observable, called an entanglement witness, which can have larger values for entangled states than separable states can achieve \cite{Horodecki1996,Terhal1999,Lewenstein2000}. When more than two particles are involved, it is common to witness that the particles are genuine multipartite entangled (GME), i.e.~that the entanglement cannot be reduced to fewer particles \cite{Guhne2009}. The entanglement witness framework typically assumes that all the lab measurements are perfectly controlled, i.e.~exactly the desired observable is measured. However, this is a strong
idealisation. Theoretically, it is known that small experimental deviations can cause large false positives  \cite{Seevinck2007,Rosset2012,MorelliYamasaki_2022}.

Imperfect measurements are, however, not a problem if one certifies entanglement device-independently, e.g. Bell inequalities  \cite{Moroder2012, Bancal2011, barreiro2013demonstration}. However, most entangled states either cannot, or are not known to, violate any Bell inequality.  Moreover, for practical purposes, relaxing the natural idea of a trusted yet imperfectly controlled device to a black-box picture is along the lines of killing a fly with an elephant gun.

To construct a more appropriate model, one may consider that the experimenter controls the measurements, but only up to a given level of precision. That is, the lab measurement only nearly corresponds to the desired measurement. Recently, for two particles, it has been shown how to construct some entanglement witnesses that take the magnitude of imperfection into account  \cite{MorelliYamasaki_2022, Tavakoli2024}. In this work, we theoretically develop and experimentally demonstrate such approaches for GME between many particles. The $n$-particle  regime presents not only a methodological obstacle but also an important conceptual question: if each single-qubit measurement can be imperfect, it appears plausible that the detrimental influence of imperfections can accumulate over the $n$ single-particle measurements, making it  hard to witness GME already for tiny imperfections. We show that this problem can be circumvented. We develop $n$-particle  GME-witnesses for the seminal Greenberger-Horne-Zeilinger (GHZ) states and show that the influence of imperfections is constant in the number of particles, thus permitting meaningful experimental use.  In a four-photon table-top experiment, we first showcase the relevance of small imperfections in the multiparticle setting by using three-photon entanglement and small measurement imperfections to emulate four-photon GME in a conventional entanglement witness test. Then, using four-photon GHZ-entanglement, we show how our methods  overcome the issue of imperfect control. In particular, we demonstrate that this can be achieved at significantly higher noise rates than in well-known  device-independent witnesses.
%

\textit{Theory.---}
In an entanglement witness test, the experimenter is asked to make a series of measurements on given basis involving each particle. For qubit systems, the target measurement corresponds to orthogonal projectors $P_{\pm}=\ketbra{\pm\vec{n}}{\pm\vec{n}}$, for some unit Bloch vector $\vec{n}$. However, due to the lack of perfect control, the lab measurements are different, $\tilde{P}_\pm$, and they can be both noisy and misaligned. Nevetheless, they closely approximate $P_\pm$.  A natural quantifier of this approximation is the average fidelity \cite{MorelliYamasaki_2022}, $    \mathcal{F}_{\vec{n}}\equiv\frac{1}{2}\bracket{\vec{n}}{\tilde{P}_+}{\vec{n}}+\frac{1}{2}\bracket{-\vec{n}}{\tilde{P}_-}{-\vec{n}}$. This can be readily estimated in the lab using an auxiliary source that prepares the eigenstates of the target measurement to probe the lab measurements. The deviation is our imprecision parameter, $\varepsilon$. We therefore permit any lab measurement which is at least $\varepsilon$-close to the target measurement, i.e.~ satisfying $ \mathcal{F}_{\vec{n}}\ge 1-\varepsilon$.  

A state comprised of $n$ subsystems is said to be biseparable if it can be generated by classical randomness, $\{q_i\}$, and quantum states, $\sigma_i$, that are separable with respect to a bisection, $i$, of the subsystems, i.e.~
$ \rho= \sum_{i} q_i \sigma_{i}$. States that are not biseparable are GME. A GME-witness is an inequality for the possible values of an observable, which is satisfied by all biseparable states but violated by some GME states.  Typically, the witness is tailored so that it performs well for states in the vicinity of the state targeted in the experiment. 

An important state in quantum information is the GHZ state, $\ket{\mathrm{ghz}_n}=\frac{1}{\sqrt{2}}\left( \ket{0}^{\otimes n} +\ket{1}^{\otimes n}\right)$  \cite{pan2000experimental,bouwmeester1999observation,gottesman1999demonstrating}. The GME of the GHZ state can be detected via the Mermin observable \cite{Mermin_1990}
\begin{align}\label{eq:Mermin}
    \mathcal{M}^{(n)} = \frac{1}{2}\bigg(\bigotimes\limits_{j=1}^n(X+iY)+\bigotimes\limits_{j=1}^n(X-iY)\bigg), 
\end{align}
where $X$, $Y$ and $Z$ denote the Pauli matrices. This corresponds to party $j$ measuring either $X$ or $Y$ on their share of the state.  A GME witness is $ \expect{\mathcal{M}^{(n)}}_\text{bisep} \leq   2^{n-2}$, which is violated by the GHZ state   because $\expect{\mathcal{M}^{(n)}}_\text{ghz}=2^{n-1}$. In addition, the GME can also be detected device-independently, i.e.~without any assumptions on the measurements,  because    $
\expect{\mathcal{M}^{(n)}}_\text{DI-bisep}\leq2^{n-3/2}$  \cite{Gisin_1998,Werner_2000,Liang_2014}.

Let us now introduce imperfect measurements for the GME-witness. The two local observables of party $j$ can now be represented as $\tilde{X}_{j}=q_{j,x}X+\sqrt{1-q_{j,x}^2}X_j^\perp$ and $\tilde{Y}_{j}=q_{j,y}Y+\sqrt{1-q_{j,y}^2}Y_j^\perp$ respectively. Here,  $X_j^\perp$ ($Y_j^\perp$) is some observable perpendicular to $X$ ($Y$) and   $1\geq q_{j,x},q_{j,y}\geq 1-2\epsilon$ correspond to the imprecision parameters. Note that perpendicular observables here means that $\tr(XX^\perp)=0$ and $\tr(YY^\perp)=0$. For simplicity, we put  $q_j=\min(q_{j,x},q_{j,y})$, which means that the largest imperfection over the two measurement is set as a standard for both settings. The Mermin operator is now no longer a constant but instead becomes $   \mathcal{M}^{(n)}_\varepsilon=
        \tfrac{1}{2}\bigotimes\limits_{j=1}^n(\tilde{X}_{j}+i\tilde{Y}_{j})+\tfrac{1}{2}\bigotimes\limits_{j=1}^n(\tilde{X}_{j}-i\tilde{Y}_{j})$.
We must determine the largest value of $\expect{\mathcal{M}^{(n)}_\varepsilon}_\rho$ under biseparable states, $\rho$, and all possible $\varepsilon$-close imperfect measurements. The latter part is therefore an optimisation over all parameters $\{q_{i}\}$ and all perpendicular observables $\{X_i^\perp\}$ and $\{Y_i^\perp\}$. The solution to this problem is 
\begin{align}\label{optboundMermin}
       \expect{\mathcal{M}^{(n)}_\varepsilon} \stackrel{\text{bisep}}{\leq}2^{n-2}\left(1-2\varepsilon+2\sqrt{\varepsilon(1-\varepsilon)}\right)
\end{align}
when $\varepsilon\leq \frac{2-\sqrt{2}}{4}$ and $\expect{\mathcal{M}^{(n)}_\varepsilon}\leq 2^{n-3/2}$ otherwise.
Moreover, the bound can be saturated by the biseparable state 
\begin{equation}
|\psi_n\rangle=\frac{1}{2}\left(|0\rangle+\mathrm{e}^{i\frac{\pi}{4}}|1\rangle \right)\otimes\left(|0\rangle^{\otimes n-1}+\mathrm{e}^{-i\frac{\pi}{4}}|1\rangle^{\otimes n-1}\right). \label{eq:tripartite entanglement}
\end{equation}
The proof is given in SM.~I.
We make three relevant comments. Firstly, in the limits of small and large imprecision parameter, we recover the standard GME-witness and the device-independent GME-witess respectively. Secondly,  in the most relevant regime, where $\varepsilon$ is small, the first-order approximation is  $\expect{\mathcal{M}^{(n)}_\varepsilon}\lesssim 2^{n-2}\left(1+2\sqrt{\varepsilon}\right)$. This shows a significant influence for small values of $\varepsilon$. Thirdly,  the $\varepsilon$-dependent correction factor in \eqref{optboundMermin} is independent of $n$. This means that witnessing GME under imperfect measurements does not come with a decreasing noise tolerance in the number of qubits. This feature is reflected in the proof since there exists a strategy that leads to saturation in \eqref{optboundMermin} in which only one party performs imperfect measurements. This  also highlights the relevance of considering imperfections; small errors already on a single party can be sufficient to reach the full potential of false positives for a standard GME witness.

Complementary to this, another well-known GME witness, tailored for the GHZ state, that only uses two settings per party, is based on stabilisers \cite{Toth1_2005,Toth_2005}. The witness operator is defined as 
\begin{align}\label{eq:stabilizer}
    \mathcal{W}^{(n)}=2^{n-2}\prod\limits_{j=1}^n X^{(j)}+\prod\limits_{j=2}^n(Z^{(j-1)} Z^{(j)}+\mathds{1})-\mathds{1},
\end{align}
where $X^{(j)}$, $Y^{(j)}$ and $Z^{(j)}$ denote the Pauli matrices on the $j$'th qubit, padded with the identity operator on all other parties. For biseparable states $
    \langle \mathcal{W}^{(n)}\rangle\leq2^{n-1}-1$, 
whereas the GHZ state achieves the algebraic maximum of $\langle \mathcal{W}^{(n)}\rangle=3\times 2^{n-2}-1$. However, in contrast to the Mermin witness, the stabilizer witness cannot be used device-independently since a local hidden variable model can reach  $3\times 2^{n-2}-1$. In analogy with before, for imperfect measurements we replace the operators $(X^{(j)},Y^{(j)},Z^{(j)})$ in \eqref{eq:stabilizer} with the imperfect observables  $(\tilde{X}^{(j)},\tilde{Y}^{(j)},\tilde{Z}^{(j)})$. As shown in SM.~II, states separable w.r.t~a partition in which both sets contain at least two qubits are bounded as  $9\times2^{n-4}-1$. This is $\varepsilon$-independent and only somewhat above the biseparable bound. Therefore, we must focus on bisecting w.r.t~a single qubit. 
W.l.g.~we can choose the first qubit; the resulting state is of the form $|\psi_n\rangle=|\chi_1\rangle|\psi_{2\dots n}\rangle$. To simplify the optimisation, note that the ideal observables are in the $X-Z$ plane. The imperfect observables can then be restricted as $\tilde{X}_{j}=q_{j}X+\sqrt{1-q_{j}^2}Z$ and $\tilde{Z}_{j}=q_{j}Z+\sqrt{1-q_{j}^2}X$. Hence, we can w.l.g.~choose $|\chi_1\rangle$ in the the $X-Z$ plane, $|\chi_1\rangle=\cos{\theta}|0\rangle+\sin{\theta}|1\rangle$. Also, we can set $q_j=1-2\epsilon$, corresponding to maximal imprecision parameters. This reduces the imperfect witness operator to depend only on $\theta$. Optimising over $\theta$ is hard analytically but can be done reliably by numerical means.  In SM. III
we analyze the case $n=3$ and $n=4$, thereby obtaining imperfection-robust witnesses.

This has several differences with the Mermin witness. For sufficiently large $\varepsilon$, the biseparable bound actually reaches the quantum bound, making low-fidelity measurements unable to detect GME. Also, while imprecise measurements in one party are enough to approximate the biseparable bound well for small values of $\varepsilon$, reaching this bound  requires maximal error in all parties. Lastly, in contrast to \eqref{eq:tripartite entanglement}, the state for which the maximal biseparable bound is reached depends on the error $\varepsilon$.

Finally, in SM.~IV we analyse also GME detection with imperfect measurements for other states than the GHZ state; specifically the three-qubit W-state and the four-qubit cluster state~\cite{Briegel_2001}.


\begin{figure}
\begin{center}
    \includegraphics[width=1\columnwidth]{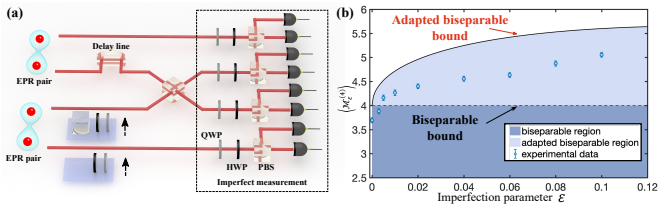}
    \caption{(a) Sketch of the experimental setup. Each EPR pair represents an entangled state $\left(|HH\rangle+|VV\rangle\right)/\sqrt{2}$ generated through spontaneous parametric down-conversion. The elements above the blue plates are inserted into optical paths to generate the four-qubit state with only tripartite entanglement. The correlations are measured through four sets of polarization analyzer setup, which consists of a quarter-wave plate (QWP), a half-wave plate (HWP), a polarisation beam splitter (PBS), and two detectors. (b) Mermin GME-witness test with misaligned measurements. The shaded areas show the biseparable regions for the standard (darker) and imperfection-adapted (lighter) witnesses. The error bars are calculated by considering the Poissonian nature of the detection events. }
    \label{fig:spoofing}
\end{center}    
\end{figure}

\textit{Experimental apparatus.---}
The basic ingredient of experiment is photonic four-partite entangled states. As depicted in Fig.~\ref{fig:spoofing} (a), the four-qubit GHZ state is generated by entangling two EPR pairs through the Hong-Ou-Mandel interference on PBS \cite{zhang2015experimental,cao2022experimental} (see SM. V for details), where the horizontal (vertical) polarization encodes the logic 0 (1).

To quantify the deviation of our experimental measurements from idealized ones, we perform a measurement tomography on the projective basis of relevant observables $X$, $Y$ and $Z$ for witnesses.
This yields fidelities of $\mathcal{F}_X=99.94\pm0.002\%$, $\mathcal{F}_Y=99.77\pm0.005\%$ and $\mathcal{F}_Z=99.97\pm0.002\%$ respectively.
All photons adopt identical measurement configurations. We note that the imprecision parameters alternatively can be estimated from simulating the systematic errors of the experiment. In SM. VI we perform such an analysis and find results consistent with the above measured values.

\textit{False positives from small alignment errors.---} We demonstrate the impact of small systematic errors on the measurement by showing false positives for the standard Mermin witness for the GME of four qubits. Our false positives are based on preparing the state \eqref{eq:tripartite entanglement} which only has three genuinely entangled qubits. As seen in Fig. \ref{fig:spoofing}, it can be realized by removing the entanglement of the second source with an additional PBS, followed by local unitaries to embed the desired phase (see SM for details of state preparation).
Then, we artificially introduce a systematic misalignment to our measurements observables of the form $\widetilde{X}=qX+\sqrt{1-q^2}Y$ and $\widetilde{Y}=qY+\sqrt{1-q^2}X$, where $q=1-2\varepsilon$, instead of $X$ and $Y$.
The optimal strategy for creating a false positive requires us to apply the misaligned observables $\widetilde{X}$ and $ \widetilde{Y}$ on the first qubit, while performing the exact observables, $X$ and $Y$, on the other three respective qubits. 
For simplicity, we will assume that all of of our measurements ($X$, $Y$, $\widetilde{X}$, and $\widetilde{Y}$) are the target measurements.
This is a reasonable assumption, as we will first study a regime wherein the introduced errors on $\widetilde{X}$ and $\widetilde{Y}$ are much larger than errors in nominal measurements of $X$ and $Y$ (which have fidelities above $99.7\%$, discussed above).

We evaluate the Mermin GME-witness for the imprecision parameters $\varepsilon\in[0,0.1]$; the results are shown in Fig.~\ref{fig:spoofing}. In particular, for the faithful implementation  ($\varepsilon=0$), we obtain $\expect{\mathcal{M}^{(4)}_0}= 3.700\pm0.053$ but this increases to $\expect{\mathcal{M}^{(4)}_\varepsilon}= 4.178\pm0.054$ already at $\varepsilon=0.5\%$,  corresponding to a $13$ percent increase and a false positive of four-qubit GME. In general,  it is expected that the amount of imprecision required for a false positive will decrease as the quality of the state preparation improves, i.e.~high-quality sources are the most vulnerable since they can come close to the idealised biseparable limit. In contrast, the solid line in Fig~\ref{fig:spoofing}  indicates our imprecision-robust GME-witness \eqref{optboundMermin}. We clearly see that all data points satisfy the witness inequality.

\textit{GME from imperfect measurements.---} We now use our imprecision-corrected GME-witnesses based on the Mermin quantity \eqref{eq:Mermin} and the stabiliser quantity \eqref{eq:stabilizer} to detect the GME of the four-qubit GHZ state.  Our experimental imprecision parameters are based on the previously reported fidelities, namely  $\varepsilon_j=1-\mathcal{F}_j, j=X, Y, Z$.  We measure the value  $\expect{ \mathcal{M}^{(4)}_\varepsilon}=7.4665\pm0.0212$ which violates the $\varepsilon$-dependent biseparable bound $\expect{\mathcal{M}^{(4)}_{\varepsilon}} \leq 4.38$, which is obtained from  Eq.~\eqref{optboundMermin} by making the worst-case choice  $\varepsilon=\mathrm{max}\left\{\varepsilon_X, \varepsilon_Y\right\}=2.5\times 10^{-3}$. 
Similarly, the evaluation of stabilizer witness gives $\expect{ \mathcal{W}^{(4)}_\varepsilon}=10.5168 \pm 0.0412$, which violates the $\varepsilon$-dependent biseparable bound $\expect{\mathcal{W}^{(4)}_{\varepsilon}} \leq 7.19$ calculated in SM. III by setting $\varepsilon=\mathrm{max}\left\{\varepsilon_X, \varepsilon_Z\right\}=6\times 10^{-4}$. 
It is worth noticing the GME in our experiments remains detected even when chosing  much larger imprecision parameters. 

\begin{figure}
    \centering
    \includegraphics[width=0.9\columnwidth]{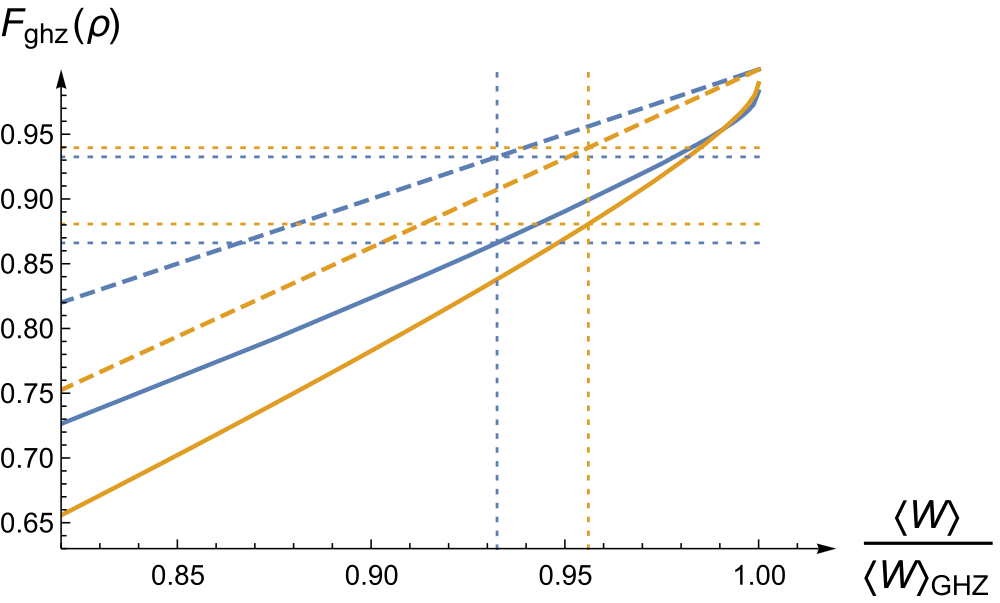}
    \caption{\textbf{Fidelity estimation with imperfect measurements.} Numerical lower bound on the GHZ-fidelity of a four-qubit state $F_{\text{ghz}}(\rho)$ versus the fraction of the optimal value obtained by a GHZ-state $\frac{\expect{W}}{\expect{W}_{GHZ}}$.
     The dotted vertical lines show the values $0.9325$ and $0.9561$ obtained in the experiment and the dotted horizontal lines show the fidelity estimation for the corresponding treatment.}
    \label{fig:fidelity}
\end{figure}

Another important feature of these GME-witnesses is that they can be used to bound the GHZ-fidelity of the state, namely $F_{\text{ghz}}(\rho)\geq  \frac{\expect{\mathcal{M}^{(4)}_0}}{8}\equiv L^{\mathcal{M}}_0$ and $F_{\text{ghz}}(\rho)\geq  \frac{\expect{\mathcal{W}^{(4)}_0-3}}{8}\equiv L^{\mathcal{W}}_0$ respectively. However, as we now show, if the measurements are imprecise, the standard fidelity bounds can be significant over-estimations. In general, we must compute imprecision-adapted lower bounds $L^{\mathcal{M}}_{\varepsilon}$ and $L^{\mathcal{W}}_{\varepsilon}$ respectively.

We have numerically searched for the smallest value of $F_{\text{ghz}}$ compatible with a given witness value $\beta$ for the experimentally observed imprecision parameters $\{\varepsilon_X,\varepsilon_Y,\varepsilon_Z\}$. That is
\begin{equation}
F_{\text{ghz}}(\rho)\geq \mathop{\min}\limits_{\expect{R^{(4)}_\varepsilon}_\sigma=\beta} F(\sigma,|\mathrm{ghz}_4\rangle)\equiv L^{R}_{\varepsilon},
\end{equation}
for $R\in\{\mathcal{M},\mathcal{W}\}$. The fidelity bound is illustrated in Fig.~\ref{fig:fidelity}. For comparison with standard witnesses, $L_{0}$ (dashed lines) and the numerically estimated $L_{\varepsilon}$ (solid lines) are plotted versus the observed fraction of the maximal value, with the Mermin and stabilizer witness denoted by blue and orange curves respectively. Despite the tiny experimental imprecision, the estimated fidelity still suffers a significant decrease compared with the assumed ideal measurement. 
Specifically, from our data, by assuming ideal measurements a fidelity of $93.25\%$ ($93.96\%$) for the Mermin (stabilizer) witness can be claimed; however, given our measurement imperfections we can only claim a fidelity of $86.61\%$ and $88.07\%$ respectively. This highlights the potential quantitative relevance of these systematic errors.

\textit{Robustness hierarchy.---}
The imperfect measurement framework is an attempt at modelling realistic measurement errors in an operational way, which requires minimal modelling of the device, without resorting to a black-box picture. Consequently, the deductions are made with much more knowledge in hand than from device-independent inference. This suggests that the decreasingly powerful frameworks, namely device-independence, imperfect measurements and idealised measurements should, respectively, be increasingly successful at detecting weakly entangled states. We now study this intuition for the most relevant form of noise in our setup. We verify that the imperfect measurement framework permits the detection of states that are device-independently not detected.

\begin{figure}[t!]
\begin{center}
    \includegraphics[width=0.95\columnwidth]{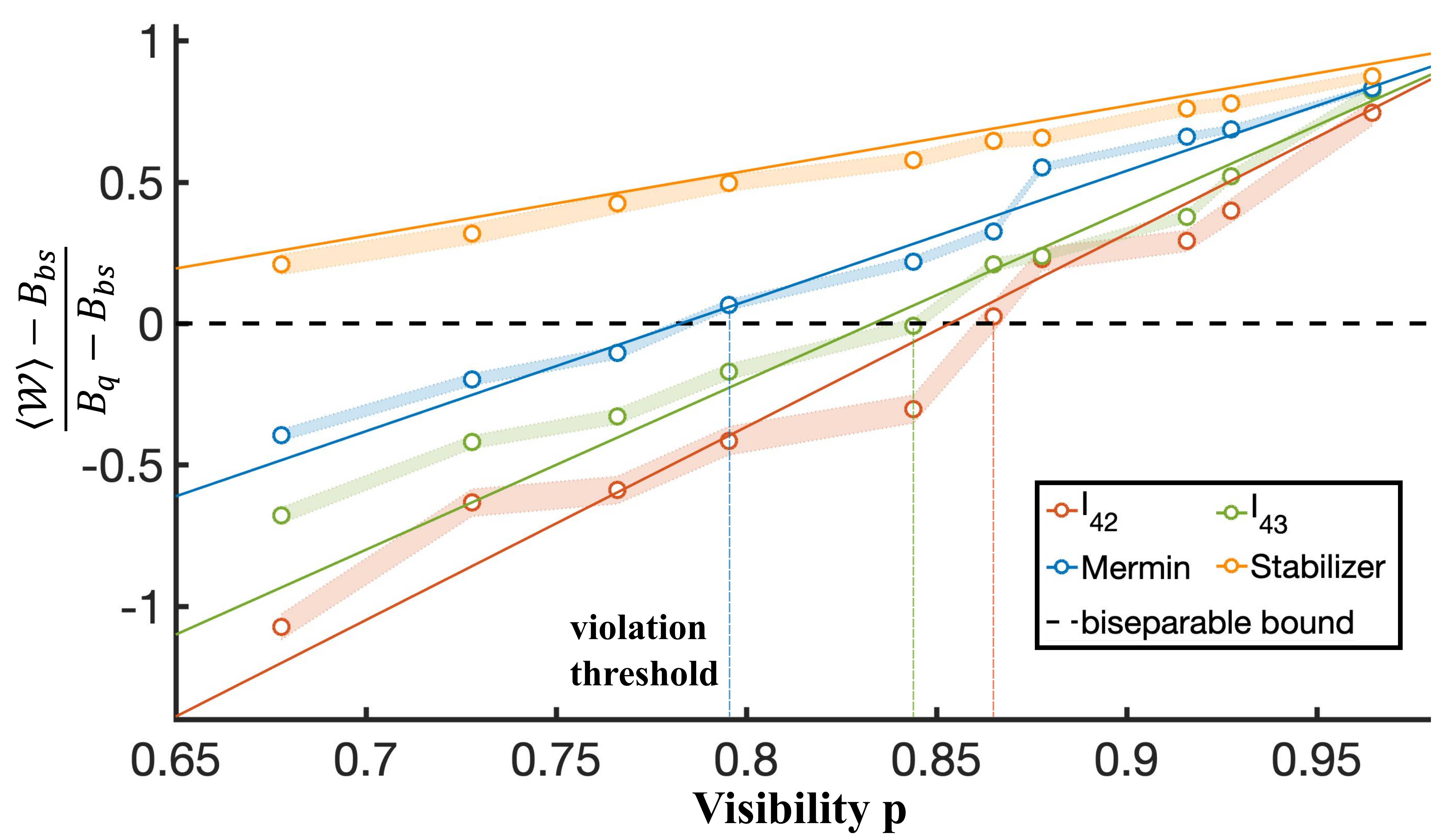}
    \caption{(b) Experimental robustness comparison between device-independent witnesses and imprecision-adapted witnesses. The horizontal axis represents the the visibility $p$ in Eq.~\eqref{eq:GHZmodel}. 
    Experimental data (theoretical prediction) are plotted by dots (solid lines) while errorbars are described with shadows around. For the convenience of comparison, we normalize the witness value $\langle \mathcal{W}\rangle$ into $\frac{\langle \mathcal{W}\rangle-B_{bs}}{B_q-B_{bs}}$, thus the biseparable bound is always 0 and the maximal violation allowed by quantum theory is always 1. $B_{bs}$: biseparable bound; $B_{q}$: quantum bound. }
    \label{fig:robustness}
\end{center}    
\end{figure}

Dephasing is a dominant type of noise in many quantum devices. We investigate a dephased GHZ  
\begin{equation}
    \rho_{\text{deph}}(p)=p\ketbra{\text{ghz}_4}
    +(1-p)\ketbra{\text{ghz}^-},\label{eq:GHZmodel}
\end{equation}
where the $|\text{ghz}^-\rangle=(|0\rangle^{\otimes 4}-|1\rangle^{\otimes 4})/\sqrt{2}$, and $p\in[0.5,1]$ quantifies the quality of the source. We use a delay line to tune the arrival time of one of the interfering photons, see Fig.~\ref{fig:spoofing}(a). This enables us to vary the strength of the dephasing. We compare our approach with a device-independent witness of GME, a witness $I_{nm}$ for $n$-partite quantum system with $m$ possible choices of local measurement settings, that was introduced in \cite{Bancal2011, Bancal_2012, barreiro2013demonstration} and experimentally demonstrated in ref.~\cite{barreiro2013demonstration} (a review on the witness is given in SM. VII). We investigate the $I_{42}$, $I_{43}$, $\mathcal{M}^{(4)}_\varepsilon$ and $\mathcal{W}^{(4)}_\varepsilon$ for varied visibility $p$ and identify the point at which the threshold of violating biseparable bound is exceeded. To showcase the relevance of our argument beyond the relatively small imperfections reported for our setup, we conservatively choose a significantly larger imprecision parameter $\varepsilon=0.5\%$. In Fig~\ref{fig:robustness}, we find the threshold point for $m=2$ (red-colored points) and $m=3$ (green-colored points) of $I_{nm}$ to be around $p=86.4\%$ and $p=84.3\%$ respectively, which is marked with vertical dashed lines with corresponding colors. Although more settings per party may slightly increase noise tolerance \cite{barreiro2013demonstration}, these are both notably lower amounts of noise than we found for the revised Mermin witness (blue-colored data), which can detect GME down to values of $p=79.4\%$. the samples of stabilizer witness are all well above the biseparable bound. See SM. VIII for extended discussions. This illustrates how the device-independent approach while excluding concerns of device control, comes at an additional price in terms of the detectable GME. By exploiting easily available knowledge about the apparatus, here the imprecision parameter, the utility of the device is significantly improved, while still doing away with the unwarranted idealisations standard entanglement witnesses.

\textit{Conclusions.---} Verifying GME is an important building block for quantum computers and quantum networks. Here, we have shown how such verification can be impacted by the inevitable presence of small systematic errors. While our experiment focuses on photonics, its relevance extends also to other platforms; for example in atomic qubits \cite{wang2016single,barreiro2013demonstration,debnath2016demonstration,fang2022crosstalk} and super-conducting qubits \cite{cao2023generation,corcoles2015demonstration,ansmann2009violation}  where cross-talk in measurement is relevant. Another frontier where measurement imperfections are particularly relevant is in high-dimensional quantum systems \cite{bent2015experimental,mirhosseini2013efficient}, where it is often the case that control over the measurement decreases with the dimensionality. 


Furthermore, one should note that there are many ways of modeling imperfect measurements. Our approach here, which builds on a series of related previous works \cite{Rosset2012, MorelliYamasaki_2022, Tavakoli2024}, has the advantage that it is operationally meaningful, does not require any detailed physical modeling, is not limited to a specific physical platform and requires a small resource cost for its estimation. However, for specific  systems, one can also make relevant models based on their specific physics, still without requiring extensive noise tomography.  An alternative route to reduce the necessary correction terms by increasing the resources spent in probing the measurement devices, i.e.~to use additional trusted parameters. All these avenues  represent relevant middle-ways between standard entanglement witnessing and device-independent entanglement certification.

\textit{Acknowledgement.---}
S.M. is supported by the Basque Government through IKUR strategy and through the BERC 2022-2025 program and by the Ministry of Science and Innovation: BCAM Severo Ochoa accreditation CEX2021-001142-S / MICIN / AEI / 10.13039/501100011033. A.T. is supported by the Wenner-Gren Foundation and by the Knut and Alice Wallenberg Foundation through the Wallenberg Center for Quantum Technology (WACQT). P.W., L.R and H.C are supported by 
the European Union’s Horizon 2020 research and innovation programme under grant agreement No 820474 (UNIQORN) and No 899368) (EPIQUS);  the Austrian Science Fund (FWF) through [F7113] (BeyondC), and [FG5] (Research Group 5); the AFOSR via FA9550-21- 1-0355 (QTRUST); the QuantERA II Programme under Grant Agreement No 101017733 (PhoMemtor); the Austrian Federal Ministry for Digital and Economic Affairs, the National Foundation for Research, Technology and Development and the Christian Doppler Research Association. C.Z. is supported by the Fundamental Research Funds for the Central Universities (Nos. WK2030000061, YD2030002015), the National Natural Science Foundation of China (No. 62075208).

\bibliography{GME.bib}

\clearpage
\newpage
\hypertarget{sec:appendix}
\appendix
\newpage
    \renewcommand{\thesubsubsection}{\Alph{section}.\arabic{subsection}.\Roman{subsubsection}}
    \renewcommand{\thesubsection}{\Alph{section}.\arabic{subsection}}
    \renewcommand{\thesection}{\Alph{section}}
    \setcounter{equation}{0}
    \numberwithin{equation}{section}
    \setcounter{figure}{0}
    \renewcommand{\thefigure}{A.\arabic{figure}}

\section{Proof of the theorem}\label{app:proof_Mermin}

Let $A_0^{(k)}$ and $A_1^{(k)}$ be two dichotomic observables for party $k$ padded with the identity operator on all other parties and define the following observables recursively
\begin{align}
    M_1 &=A_0^{(1)}\\
    N_1 &=A_1^{(1)}\\
    M_k &=M_{k-1}A_0^{(k)}-N_{k-1}A_1^{(k)}\\
    N_k &=M_{k-1}A_1^{(k)}+N_{k-1}A_0^{(k)}.
\end{align}
Then it holds that
\begin{align}
    M_k &= \tfrac{1}{2}\bigg(\prod\limits_{j=1}^k(A_0^{(j)}+iA_1^{(j)})+\prod\limits_{j=1}^k(A_0^{(j)}-iA_1^{(j)})\bigg)\\
    N_k &= \tfrac{-i}{2}\bigg(\prod\limits_{j=1}^k(A_0^{(j)}+iA_1^{(j)})-\prod\limits_{j=1}^k(A_0^{(j)}-iA_1^{(j)})\bigg).
\end{align}
It certainly holds for $k=2$ and therefore we can inductively conclude
\begin{align*}
    M_k =& M_{k-1}A_0^{(k)}-N_{k-1}A_1^{(k)}\\
    =&\tfrac{1}{2}\bigg(\prod\limits_{j=1}^{k-1}(A_0^{(j)}+iA_1^{(j)})+\prod\limits_{j=1}^{k-1}(A_0^{(j)}-iA_1^{(j)})\bigg)A_0^{(k)}\\
    &+\tfrac{i}{2}\bigg(\prod\limits_{j=1}^{k-1}(A_0^{(j)}+iA_1^{(j)})-\prod\limits_{j=1}^{k-1}(A_0^{(j)}-iA_1^{(j)})\bigg)A_1^{(k)}\\
    =&\tfrac{1}{2}\bigg(\prod\limits_{j=1}^{k-1}(A_0^{(j)}+iA_1^{(j)})(A_0^{(k)}+iA_1^{(k)})\\
    &+\prod\limits_{j=1}^{k-1}(A_0^{(j)}-iA_1^{(j)})(A_0^{(k)}-iA_1^{(k)})\bigg)\\
    =& \tfrac{1}{2}\bigg(\prod\limits_{j=1}^k(A_0^{(j)}+iA_1^{(j)})+\prod\limits_{j=1}^k(A_0^{(j)}-iA_1^{(j)})\bigg)
\end{align*}
and analogously for $N_k$.

It further holds that
\begin{align}\label{app:eq:split}
    M_k = M_{k-l}M_l-N_{k-l}N_l,
\end{align}
up to proper relabelling of the parties, where $M_{k-l}$ and $N_{k-l}$ act on party 1 to $k-l$ and $M_{l}$ and $N_{l}$ act on party $k-l+1$ to $k$.
For $l=1$ this follows directly from the definition, so the statement follows inductively
\begin{align*}
    M_k =&M_{k-l}M_l-N_{k-l}N_l\\
    =&M_{k-l-1}A_0M_l-N_{k-l-1}A_1M_l\notag\\
    &-M_{k-l-1}A_1N_l-N_{k-l-1}A_0N_l\\
    =&M_{k-l-1}(A_0M_l-A_1N_l)-N_{k-l-1}(A_1M_l+A_0N_l)\\
    =&M_{k-l-1}M_{l+1}-N_{k-l-1}N_{l+1}
\end{align*}
after proper relabelling of the party numbers.

Following Uffink~\cite{Uffink_2002}, who proved this result for $k=2$, we show that for $k\ge2$
\begin{align}\label{app:eq:quadratic_Mermin}
    \expect{M_k}^2+\expect{N_k}^2\le 2^{2k-2}.
\end{align}

This follows from

\begin{align*}
    &\sup(\expect{M_k}^2+\expect{N_k}^2)\\
    =&\sup(\expect{M_{k-1}A_0^{(k)}-N_{k-1}A_1^{(k)}}^2+\expect{M_{k-1}A_1^{(k)}+N_{k-1}A_0^{(k)}}^2)\\
    =&\sup(\expect{M_{k-1}A_0^{(k)}}^2+\expect{N_{k-1}A_1^{(k)}}^2-2\expect{M_{k-1}A_0^{(k)}}\expect{N_{k-1}A_1^{(k)}}\\
    &+\expect{M_{k-1}A_1^{(k)}}^2+\expect{N_{k-1}A_0^{(k)}}^2+2\expect{M_{k-1}A_1^{(k)}}\expect{N_{k-1}A_0^{(k)}})\\
    \le&\sup(\expect{M_{k-1}A_0^{(k)}}^2+\expect{N_{k-1}A_1^{(k)}}^2+\expect{M_{k-1}A_1^{(k)}}^2+\expect{N_{k-1}A_0^{(k)}}^2)\\
    &+2\sup(\expect{M_{k-1}A_1^{(k)}}\expect{N_{k-1}A_0^{(k)}}-\expect{M_{k-1}A_0^{(k)}}\expect{N_{k-1}A_1^{(k)}})\\
    \le&2\sup(\expect{M_{k-1}}^2+\expect{N_{k-1}}^2)+4\sup(|\expect{M_{k-1}}\expect{N_{k-1}}|)\\
    \le&4\sup(\expect{M_{k-1}}^2+\expect{N_{k-1}}^2)\\
    \le& 2^{2k-2}.
\end{align*}
Uffink~\cite{Uffink_2002} proved a similar result in the same way, that coincides with this one for $k$ odd.

Since the previous inequality holds for all points on a disc, also every tangential inequality of the form
\begin{align}\label{app:eq:linear_Mermin}
    \alpha\expect{M_k}+\beta\expect{N_k}\le 2^{k-1}\sqrt{\alpha^2+\beta^2}
\end{align}
holds for all $\alpha$ and $\beta$.

Since the Mermin witness in Eq. (9) 
is linear, its maximal value is attained for pure states.
First, assume that we have a biseparable $n$-qubit state where each partition has more than one qubit.
Then from Eq.~(\ref{app:eq:split}) it follows that
\begin{align}
    \expect{M_n}&=\expect{M_{n-l}}\expect{M_l}-\expect{N_{n-l}}\expect{N_l}\\
    &\le 2^{n-l-1}\sqrt{\expect{M_l}^2+\expect{N_l}^2}\\
    &\le 2^{n-l-1}2^{l-1}=2^{n-2},
\end{align}
where we have first used Eq.~(\ref{app:eq:linear_Mermin}) and then Eq.~(\ref{app:eq:quadratic_Mermin}).
This is a remarkable result, it shows that the Mermin witness is completely robust against misalignment of the measurement direction, as long as the biseparable partition includes more than a single qubit in each set.

Next, we assume that only one qubit, without loss of generality, say the $n$-th qubit, is separable from the rest, such that
\begin{align}\label{app:eq:one_qubit_separable}
    \expect{M_n}&=\expect{A_0^{(n)}}\expect{M_{n-1}}-\expect{A_1^{(n)}}\expect{N_{n-1}}\notag\\
    &\le 2^{n-2}\sqrt{\expect{A_0^{(n)}}^2+\expect{A_1^{(n)}}^2}.
\end{align}

To maximise the inequality from Eq.~(\ref{app:eq:one_qubit_separable}) while satisfying the condition $A_0^{(n)}=qX+\sqrt{1-q^2}X^\perp$ and $A_1^{(n)}=qY+\sqrt{1-q^2}Y^\perp$ with $q\ge 1-2\varepsilon$, we can choose $X^\perp=Y$ and $X^\perp=Y$. Therefore it follows

\begin{align}
    &\sqrt{\expect{A_0^{(n)}}^2+\expect{A_1^{(n)}}^2}\\
    =&\sqrt{\expect{X}^2+\expect{Y}^2+4q\sqrt{1-q^2}\expect{X}\expect{Y}}\\
    \le&\sqrt{1+2q\sqrt{1-q^2}}\\
    =&(q+\sqrt{1-q^2})\\
    \le&1-2\varepsilon+2\sqrt{\varepsilon(1-\varepsilon)}
\end{align}
for $\varepsilon\le(2-\sqrt{2})/4$, where we have used $\expect{X}^2+\expect{Y}^2\le1$ and $\expect{X}\expect{Y}\le1/2$.
This proves the bound given in Eq. (10).

Assume now that party 1 performs the measurements $A_0^{(1)}=(1-2\varepsilon)X+2\sqrt{\varepsilon(1-\varepsilon)}Y$ and $A_1^{(1)}=(1-2\varepsilon)Y+2\sqrt{\varepsilon(1-\varepsilon)}X$, while all other parties perform the measurements $A_0^{(k)}=X$ and $A_1^{(k)}=Y$.
In this case the state defined in Eq. (11)
saturates the bound.


\section{Analysis of the biseparable bound for the stabilizer witness}\label{app:analysis_stabilizer}

Define the operators
\begin{align}
    M_k &=2^{k-2}\prod\limits_{j=1}^kA_0^{(j)}\\
    N_k &=\prod\limits_{j=2}^k(A_1^{(j-i)}A_1^{(j)}-\mathds{1}),
\end{align}
where again $A_0^{(j)}$ and $A_1^{(j)}$ are two dichotomic observables for party $j$ padded with the identity operator on all other parties.
It is clear that for $k\ge2$ it holds
\begin{align}\label{app:eq:quadratic_stabilizer}
    \expect{M_k}^2+c^2\expect{N_k}^2\le 2^{2k-4}+2^{2k-2}=2^{2k-4}(4c^2+1).
\end{align}
As before, also every tangential inequality of the form
\begin{align}\label{app:eq:linear_stabilizer}
    \alpha\expect{M_k}+\beta c\expect{N_k}\le 2^{k-2}\sqrt{4c^2+1}\sqrt{\alpha^2+\beta^2}
\end{align}
holds for all $\alpha$ and $\beta$.
Assume now that we have a biseparable pure $n$-qubit state, where each partition has more than a single qubit. Then it follows that 
\begin{align}
    \expect{M_n}+\expect{N_n}&\le\expect{M_{n-l}}\expect{M_l}+2\expect{N_{n-l}}\expect{N_{l}}\\
    &\le 3\times2^{n-l-3}\sqrt{\expect{M_l}^2+2\expect{N_l}^2}\\
    &\le 9\times2^{n-4},
\end{align}
where we have first used Eq.~(\ref{app:eq:linear_stabilizer}) and then Eq.~(\ref{app:eq:quadratic_stabilizer}), both with $c=\sqrt{2}$.
Since the stabilizer witness with measurement operators $A_0^{(j)}$ and $A_1^{(j)}$ is just
\begin{align}
    \mathcal{W}^{(n)}=M_n+N_n-\mathds{1},
\end{align}
it follows that the expectation value for a biseparable pure state with each partition including at least 2 qubits can not surpass the value $9\times2^{n-4}-1$, independent of the assumed error in the measurements. This value is just slightly larger than the biseparable bound for idealised measurements, see Fig.~\ref{fig:stab3} and~\ref{fig:stab4}. Therefore we focus on biseparable states where only a single party is separable from the rest.


\section{The stabilizer witness for $n=3$ and $n=4$}\label{app:stabilizer}

Here we explicitely calculate the biseparable bound for the stabiliizer witness with imprecise measurements, that is, we compute the maximal expectation value of biseparable states defined in Eq. (12).
This would in principle involve a optimization over all parameters $q_j\ge1-2\varepsilon$ and all perpendicular observables $\{X_j^\perp\}$ and $\{Y_j^\perp\}$.
Following the discussion in Section IV B of the main text
, we can reduce the problem by assuming $\tilde{X}_{j}=q_{j}X+\sqrt{1-q_{j}^2}Z$, $\tilde{Z}_{j}=q_{j}Z+\sqrt{1-q_{j}^2}X$ and $q_j=1-2\varepsilon$ for all $j$. Further we assume that the state has the form $|\psi\rangle=|\chi_A\rangle|\psi_{BC}\rangle$ with $|\chi(\theta)\rangle=\cos{\theta}|0\rangle+\sin{\theta}|1\rangle$.
This then results in
\begin{align}
    \langle\chi(\theta)|\tilde{X}_{1}|\chi(\theta)\rangle&=2\sqrt{\varepsilon(1-\varepsilon)}\cos{2\theta}+(1-2\varepsilon)\sin{2\theta}\label{app::eq:expectation_value_qubit1}\\
    \langle\chi(\theta)|\tilde{Z}_{1}|\chi(\theta)\rangle&=(1-2\varepsilon)\cos{2\theta}+2\sqrt{\varepsilon(1-\varepsilon)}\sin{2\theta}.\label{app::eq:expectation_value_qubit2}
\end{align}

\begin{figure}[t]
     \includegraphics[width=1\columnwidth]{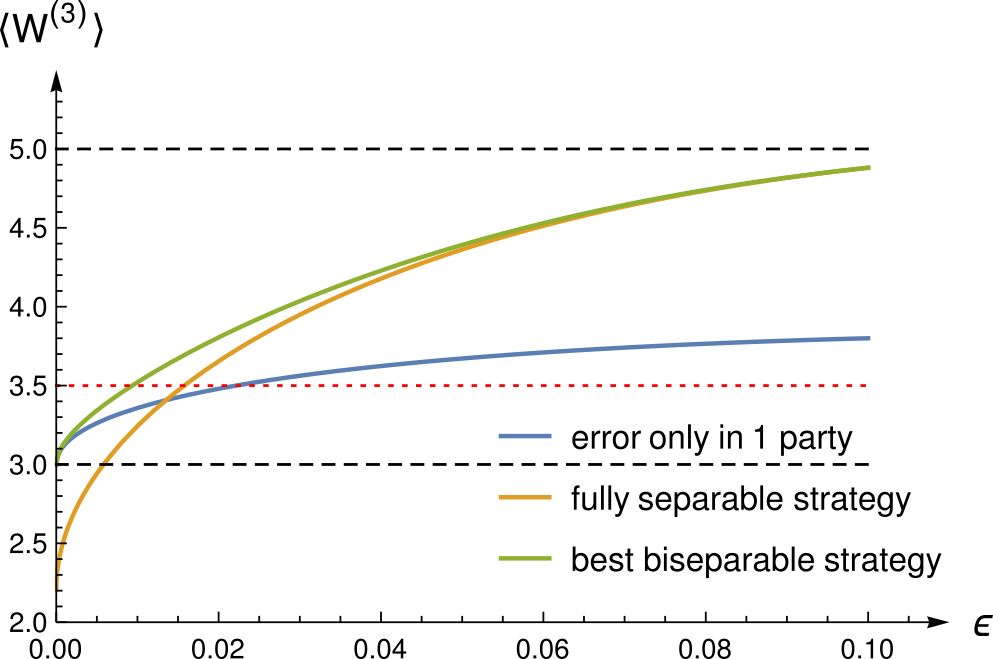}
        \caption{
        \textbf{Separable bounds of the stabilizer witness for three qubits.}
        The figure shows the separable bound of the stabilizer witness $\mathcal{W}^{(3)}$ from Eq.~(\ref{eq:stabilizer3imprecise}) with various assumptions of imprecision and separability. The biseparable bound assuming imprecise measurements in only one party is shown in blue, whereas the fully separable bound assuming imprecise measurements in all three parties is shown in orange. In green we have the maximal attainable value for biseparable states assuming imprecise measurements in all three parties.
        The dashed black lines show the biseparable and the quantum bound for accurate measurements and the dotted red line shows the bound of $9\times2^{n-4}-1$ for biseparable strategies splitting more than a single qubit.
        }
    \label{fig:stab3}
\end{figure}

For $n=3$ the stabilizer witness of Eq. (7)
with imprecise measurements becomes
\begin{align}\label{eq:stabilizer3imprecise}
    \mathcal{W}^{(3)}_\varepsilon =& 2\tilde{X}_{1}\tilde{X}_{2}\tilde{X}_{3}+\mathds{1}\tilde{Z}_{2}\tilde{Z}_{3}+\tilde{Z}_{1}\mathds{1}\tilde{Z}_{3}+\tilde{Z}_{1}\tilde{Z}_{2}\mathds{1}.
\end{align}
We therefore compute the maximal eigenvalue of
\begin{align}
    2\langle\tilde{X}_{1}\rangle \tilde{X}_{2}\tilde{X}_{3}+\langle\tilde{Z}_{1}\rangle(\mathds{1}\tilde{Z}_{3}+\tilde{Z}_{2}\mathds{1})+\tilde{Z}_{2}\tilde{Z}_{3},
\end{align}
and optimise over $\theta$.
The result, which admits no closed expression, is shown as the green line in Fig.~\ref{fig:stab3}.

Assuming errors in the measurements are only present in one party, the biseparable bound is given by the largest eigenvalue of 
\begin{align}
    \langle\tilde{X}_{1}\rangle XX+\langle\tilde{Z}_{1}\rangle(\mathds{1}Z+Z\mathds{1}) + ZZ.
\end{align}
Optimizing over $\theta$ we obtain $1+2\sqrt{1+4(1-2\varepsilon)\sqrt{\varepsilon(1-\varepsilon)}})$, shown in blue in Fig.~\ref{fig:stab3}.
We notice that for small errors $\epsilon$ this curve has the same scaling as the optimal bound.

Assuming a fully separable state $|\psi\rangle=|\chi(\theta)\rangle|\chi(\theta)\rangle|\chi(\theta)\rangle$ and $\theta=\pi/8$ we have
\begin{align}
    \langle W_{\varepsilon}^{(3)}\rangle=& \frac{3}{2}(1+\sqrt{2})-3\sqrt{2}\varepsilon -\sqrt{2}(1-2\varepsilon)^3\\
    &+2(3+\frac{\sqrt{2}}{2}-6\varepsilon+\sqrt{2}(1-2\varepsilon)^2)\sqrt{\varepsilon(1-\varepsilon)},\notag
\end{align}
which is the maximal value achievable for fully separable states. It is shown in yellow in Fig.~\ref{fig:stab3} and it approximates the optimal bound for large errors close to  $\epsilon=\tfrac{2-\sqrt{2}}{4}$.

\begin{figure}[t]
     \includegraphics[width=1\columnwidth]{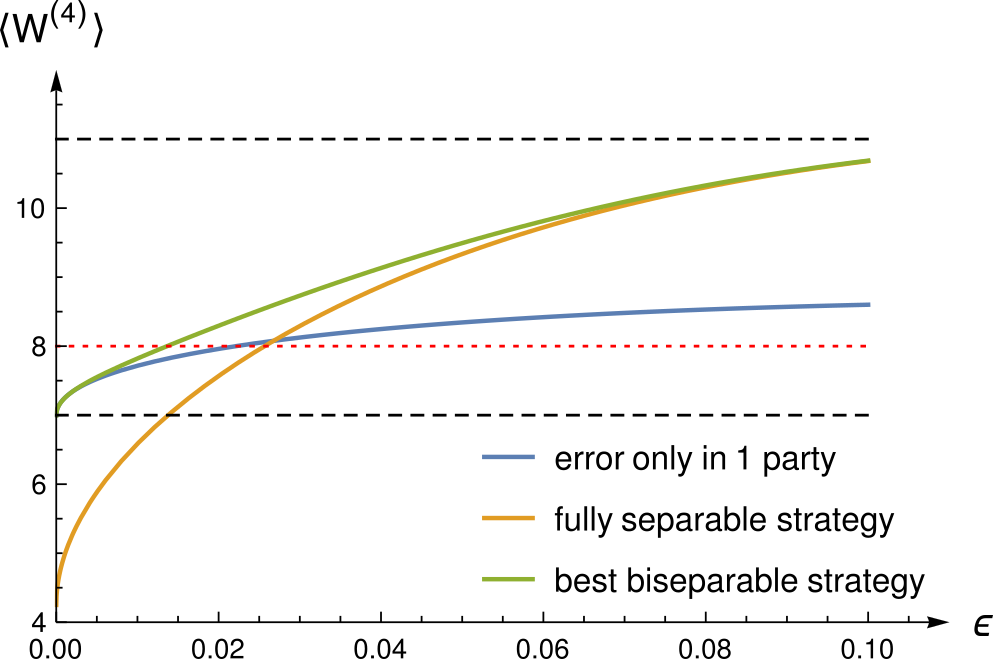}
        \caption{
        \textbf{Separable bounds of the stabilizer witness for four qubits.}
        The figure shows the separable bound of the stabilizer witness $\mathcal{W}^{(4)}$ from Eq.~(\ref{eq:stabilizer4imprecise}) with various assumptions of imprecision and separability. The green line is the conjectured optimum. On the two extremes we have in blue the case where only the measurements in the party that is separated from the rest are altered, whereas in orange we have the bound for fully separable states. The quantum bound is shown in red, it is reached both by biseparable and fully separable states for high enough misalignments in the measurement direction.
        The dashed black lines show the biseparable and the quantum bound for accurate measurements and the dotted red line shows the bound of $9\times2^{n-4}-1$ for biseparable strategies splitting more than a single qubit.
        }
    \label{fig:stab4}
\end{figure}

For $n=4$ the witness becomes
\begin{align}\label{eq:stabilizer4imprecise}
    \mathcal{W}_{\varepsilon}^{(4)}=& 4\tilde{X}_{1}\tilde{X}_{2}\tilde{X}_{3}\tilde{X}_{4}+\mathds{1}\mathds{1}\tilde{Z}_{3}\tilde{Z}_{4}+\mathds{1}\tilde{Z}_{2}\mathds{1}\tilde{Z}_{4}\notag\\
    &+\mathds{1}\tilde{Z}_{2}\tilde{Z}_{3}\mathds{1}+\tilde{Z}_{1}\mathds{1}\mathds{1}\tilde{Z}_{4}+\tilde{Z}_{1}\mathds{1}\tilde{Z}_{3}\mathds{1}\notag\\
    &+\tilde{Z}_{1}\tilde{Z}_{2}\mathds{1}\mathds{1}+\tilde{Z}_{1}\tilde{Z}_{2}\tilde{Z}_{3}\tilde{Z}_{4}.
\end{align}
We therefore compute the maximal eigenvalue of
\begin{align}
    4\langle\tilde{X}_{1}\rangle \tilde{X}_2\tilde{X}_3\tilde{X}_4+\mathds{1}\tilde{Z}_3\tilde{Z}_4+\tilde{Z}_2\mathds{1}\tilde{Z}_4+\tilde{Z}_2\tilde{Z}_3\mathds{1}\notag\\
    +\langle\tilde{Z}_{1}\rangle(\mathds{1}\mathds{1}\tilde{Z}_4+\mathds{1}\tilde{Z}_3\mathds{1}+\tilde{Z}_2\mathds{1}\mathds{1}+\tilde{Z}_2\tilde{Z}_3\tilde{Z}_4),
\end{align}
and maximise over $\theta$.
The result, which admits no closed expression, is shown as the green line in Fig.~\ref{fig:stab4}.

Assuming that imprecisions are only present in one party, we calculate the largest eigenvalue of 
\begin{align}
    4\langle\tilde{X}_{1}\rangle XXX + \mathds{1}ZZ+Z\mathds{1}Z+ZZ\mathds{1}\\
    +\langle\tilde{Z}_{1}\rangle(\mathds{1}\mathds{1}Z+\mathds{1}Z\mathds{1}+\mathds{1}Z\mathds{1}+ZZZ),\notag
\end{align}
and optimise over $\theta$, which results in $3+4\sqrt{1+4(1-2)\varepsilon\sqrt{\varepsilon(1-\varepsilon)}}$ for $\theta=\pi/8$, shown as blue line in Fig.~\ref{fig:stab4}. This case is a good approximation to the biseparable bound for small values of $\varepsilon$.

For large misalignment errors $\varepsilon$, we find that the expectation value of the fully separable state $|\psi\rangle=|\chi(\pi/8)\rangle|\chi(\pi/8)\rangle|\chi(\pi/8)\rangle|\chi(\pi/8)\rangle$ assuming equally imprecise measurements in all parties is
$\frac{17}{4}+20\varepsilon(1-\varepsilon)(1-2\varepsilon)^2+22(1-2\varepsilon)\sqrt{\varepsilon(1-\varepsilon)}$,
which approaches the biseparable bound for large errors close to $\epsilon=\tfrac{2-\sqrt{2}}{4}$. It is shown in yellow in Fig.~\ref{fig:stab4}.

\section{Analyzing entanglement witnesses for other entangled states}\label{app:witness_W_cluster}

In this appendix we investigate the impact of imprecisions in the measurements on various entanglement witnesses for states other than the GHZ-state. We thereby focus on witnesses based on the stabilizer formalism with two measurement settings per side.
For three qubits we investigate a witness designed to detect entangled states close to the W-state and for four qubits we investigate a witness designed to detect entangled states close to the cluster state.

\subsection{Stabilizer witness for 3-qubit W-state}\label{app:W_witness}

The expectation value of the operator
\begin{align}\label{app:eq:w_witness}
    D^{(3)}&= XX\mathds{1}+X\mathds{1}X+\mathds{1}XX+YY\mathds{1}+Y\mathds{1}Y+\mathds{1}YY
\end{align}
is bounded by $\langle D^{(3)}\rangle\le1+\sqrt{5}$ for biseparable states, whereas for the W-state we have $\langle D^{(3)}\rangle=4$~\cite{Toth_2005}. It is easy to see that the local bound is 6, which coincides with the algebraic maximum.

\begin{figure}[t]
     \includegraphics[width=1\columnwidth]{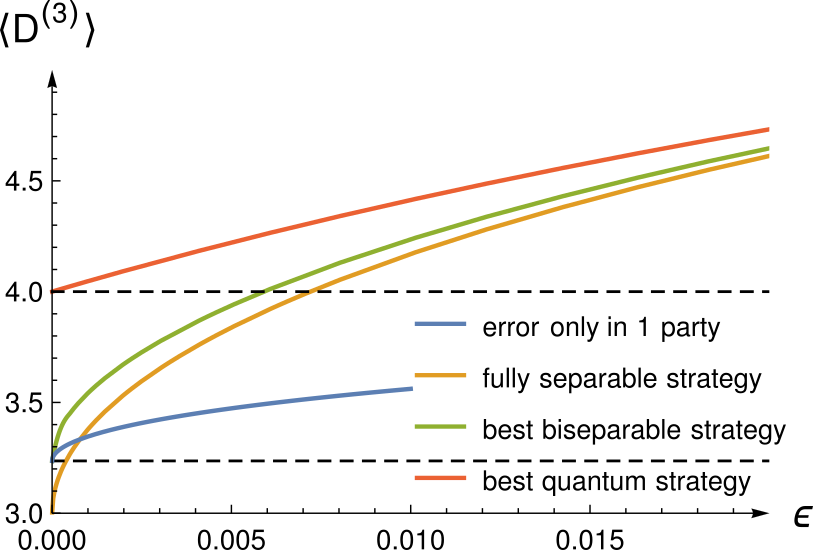}
        \caption{
        \textbf{Separable bounds for three-qubit W states.}
        The figure shows the separable bound of the stabilizer witness $D^{(3)}$ defined in Eq.~(\ref{app:eq:w_witness}) with various assumptions of imprecision and separability.
        In blue we see the bound when assuming that the error is only present in one party, whereas in yellow we see the bound on fully separable strategies. The conjectured bound on biseparable strategies is shown in green and the quantum bound is shown in orange. The dashed black lines show the biseparable and quantum bound for no error in the measurements.
        }
    \label{fig:w}
\end{figure}

To find the best biseparable strategy for imprecise measurements we can restrict to pure states and without loss of generality we assume that the first party is separable from the rest. The state thus has the form $|\psi\rangle=|\xi_A\rangle|\psi_{BC}\rangle$.
Since we are measuring in the $X-Y$ plane we further assume $|\xi\rangle=1/\sqrt{2}(|0\rangle+\mathrm{e}^{i\theta}|1\rangle)$ and noting that the remaining operator is symmetric in the two expectation values of the first party we set $\theta=\pi/4$.
This then results in the operator
\begin{align}
    \tfrac{q+\sqrt{1-q^2}}{\sqrt{2}}\left(\tilde{X}\mathds{1}+\mathds{1}\tilde{X}+\tilde{Y}\mathds{1}+\mathds{1}\tilde{Y}\right)+\tilde{X}\tilde{X}+\tilde{Y}\tilde{Y},
\end{align}
for which we calculate the maximal eigenvalue. The result is shown in green in Fig.~\ref{fig:w}.

Similar to before we can calculate the optimal biseparable strategy if imprecisions in the measurements are only present in one party and the best fully separable strategy. We find the bounds $1+\sqrt{5+16(1-2\epsilon)\sqrt{\epsilon(1-\epsilon)}}$ and $3(1+4(1-2\epsilon)\sqrt{\epsilon(1-\epsilon)})$ respectively, shown in blue and yellow in Fig.~\ref{fig:w}.

We notice a few differences to the stabilizer witness. The local bound here is larger than the quantum bound and already for an error of $\epsilon\approx0.006$ the biseparable bound surpasses the original quantum bound. Funnily, this does not imply that entanglement detection becomes impossible, as also the quantum bound increases with the error in the measurements and becomes $2 (1+\sqrt{1+48\epsilon(1-\epsilon)(1-2 \epsilon )^2 })$, shown in orange in Fig.~\ref{fig:w}

\subsection{Stabilizer witness for 4-qubit cluster state}\label{app:4_qubit_cluster1}

The expectation value of the operator
\begin{align}\label{app:eq:cluster_witness}
    C^{(4)}=& XZ\mathds{1}\mathds{1}+X\mathds{1}XZ+\mathds{1}ZXZ\notag\\
    &+ZXZ\mathds{1}+\mathds{1}\mathds{1}ZX+ZX\mathds{1}X
\end{align}
is bounded by $\langle C^{(4)}\rangle\le4$ for biseparable states, whereas for the  cluster state we have $\langle C^{(4)}\rangle=6$~\cite{Toth_2005}. This coincides both with the local bound and the algebraic maximum.

\begin{figure}[t]
     \includegraphics[width=1\columnwidth]{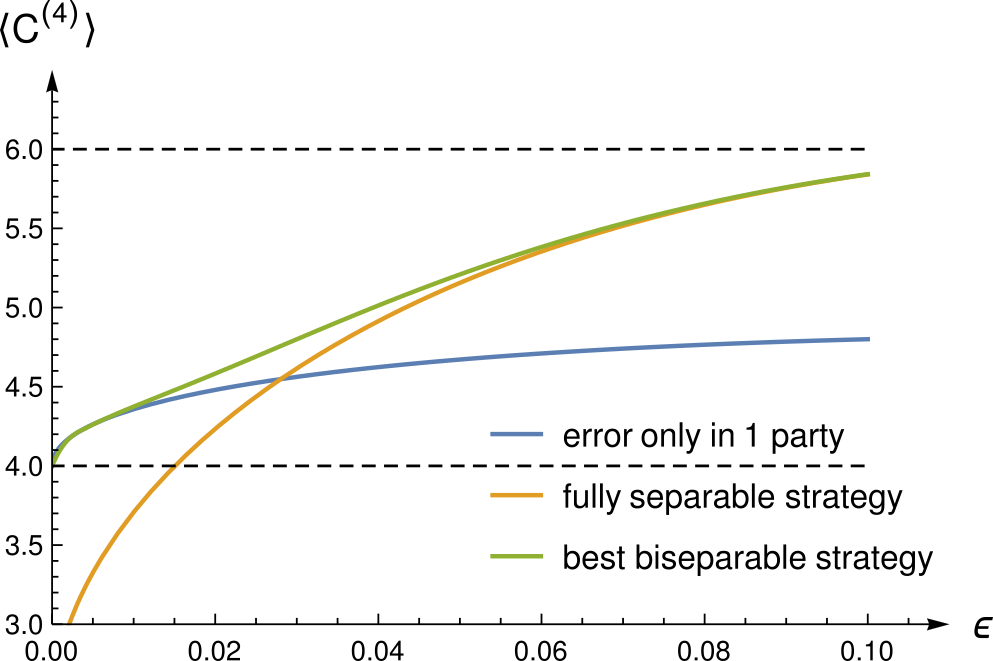}
        \caption{
        \textbf{Separable bounds for four-qubit cluster states.}
        The figure shows the separable bound of the stabilizer witness $\mathcal{C}^{(4)}$ from Eq.~(\ref{app:eq:cluster_witness}) with various assumptions of imprecision and separability. The green line is the conjectured optimum. On the two extremes we have in blue the case where only the measurements in the party that is separated from the rest are altered, whereas in orange we have the bound for fully separable states. The quantum bound is shown in red, it is reached both by biseparable and fully separable states for high enough misalignments in the measurement direction.
        The dashed black lines show the biseparable and the quantum bound for accurate measurements.
        }
    \label{fig:cluster1}
\end{figure}

We now want to find the biseparable bound for imprecise measurements.
Since this witness is linear, its maximal value is attained for a pure state. 
Assume that the first party is separable from the rest, such that our state has the form $|\psi\rangle=|\chi_A\rangle|\psi_{BCD}\rangle$.
Since we are measuring in the $X-Z$ plane we set $|\chi(\theta)\rangle=\cos{\theta}|0\rangle+\sin{\theta}|1\rangle$ as before, resulting in the same expectation value as in Eq.~(\ref{app::eq:expectation_value_qubit1})-(\ref{app::eq:expectation_value_qubit2}).
To find the optimal biseparable strategy we calculate the eigenvalue of
\begin{align}
    &\expect{\tilde{X}_1}(\tilde{Z}_2\mathds{1}\mathds{1}+\mathds{1}\tilde{X}_3\tilde{Z}_4)+\expect{\tilde{Z}_1} (\tilde{X}_2\tilde{Z}_3\mathds{1}+\tilde{X}_2\mathds{1}\tilde{X}_4)\notag\\
    &+\mathds{1}\tilde{Z}_3\tilde{X}_4+\tilde{Z}_2\tilde{X}_3\tilde{Z}_4.
\end{align}
and optimise over $\theta$. The result is shown in green in Fig.~\ref{fig:cluster1}).

Assuming errors in the measurements are only present in one party, the biseparable bound is the maximal eigenvalue of
\begin{align}
    \expect{\tilde{X}_1}(Z\mathds{1}\mathds{1}+\mathds{1}XZ)+\expect{\tilde{Z}_1} (XZ\mathds{1}+X\mathds{1}X)+\mathds{1}ZX+ZXZ,
\end{align}
optimised over $\theta$. This results in $2(1+\sqrt{1+4(1-2\epsilon)\sqrt{\epsilon(1-\epsilon)}})$ for $\theta=\pi/8$, shown in blue in Fig.~\ref{fig:cluster1}).
The optimal fully separable strategy is obtained for the state $|\psi\rangle=|\chi(\theta)\rangle|\chi(\theta)\rangle|\chi(\theta)\rangle|\chi(\theta)\rangle$ and $\theta=\pi/8$, which results in the bound $1+ 2 \sqrt{2} \sqrt{\epsilon(1-\epsilon)}+(1-2 \epsilon)\big(4 \sqrt{-(\epsilon -1) \epsilon }+3 \sqrt{2}+2 \sqrt{2} (1-2 \epsilon ) (2 \epsilon +2 \sqrt{-(\epsilon -1) \epsilon }-1)\big)$, 
shown in orange in Fig.~\ref{fig:cluster1}.

\section{Experimental details of photonic source and four-qubit state preparation}
\begin{figure}
\begin{center}
    \includegraphics[width=1\columnwidth]{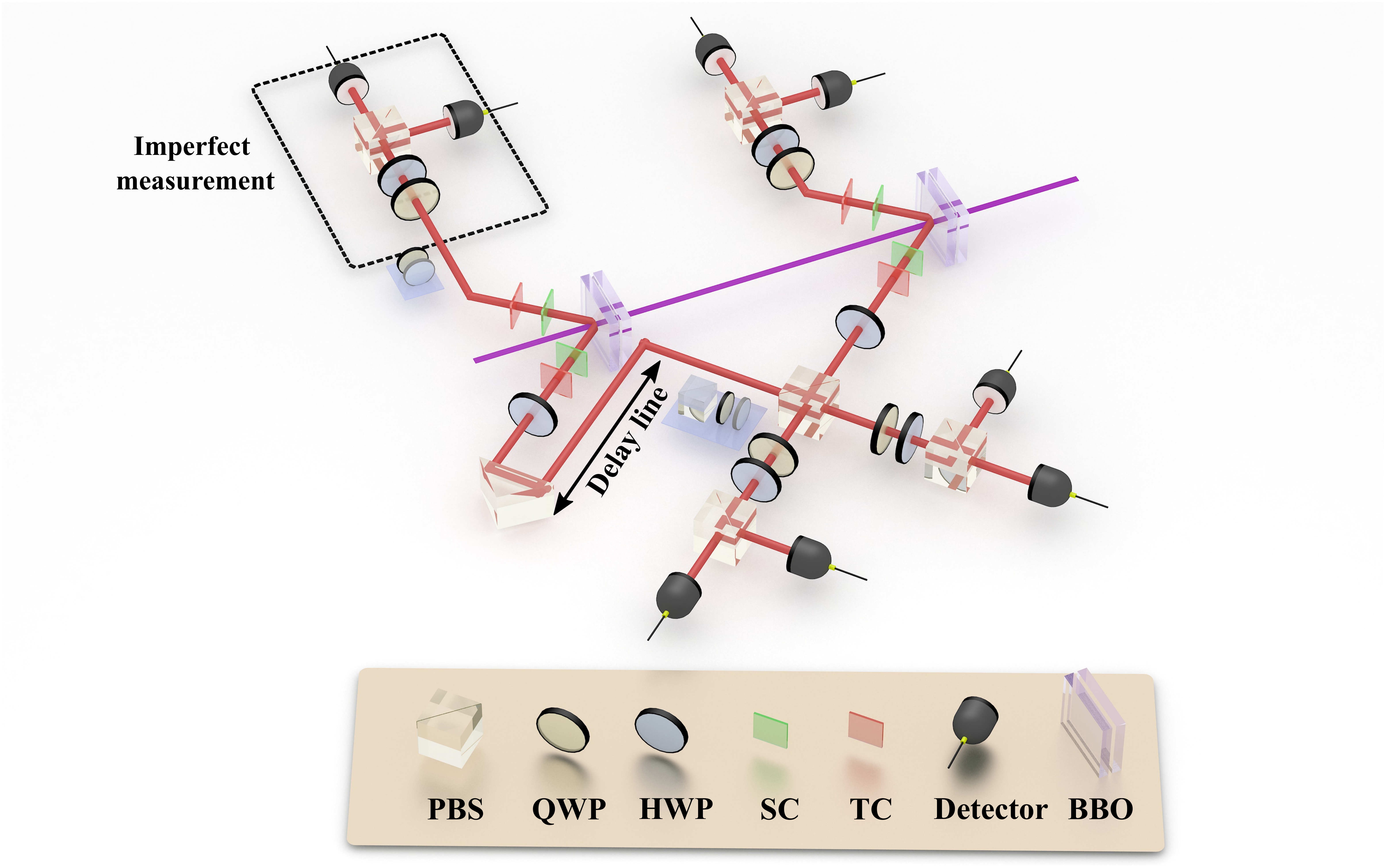}
    \caption{\textbf{Sketch of the experimental setup.}  An ultraviolet pulse pumps two cascaded sandwich-like EPR sources, generating an EPR pair $\left(|HV\rangle+|VH\rangle\right)/\sqrt{2}$ in each source. 
    The correlations are measured through four sets of polarization analyzer setup (PAS), which consists of a quarter-wave plate (QWP), a half-wave plate (HWP), a polarisation beam splitter (PBS), and two detectors.}
    \label{fig:4GHZ}
\end{center}    
\end{figure}
\textit{Sandwiched-like geometry of EPR source.--} The 4-photon entangled-state is based on two sandwich-like spontaneous parametric down-conversion (SPDC) sources \cite{zhang2015experimental,cao2022experimental}. 
The configuration is presented in Fig.~\ref{fig:4GHZ}, two cascaded sandwiched-like EPR sources are pumped by a 390 nm ultraviolet light generated via a frequency doubler with a mode-locked Ti: sapphire femtosecond laser of 780 nm central wavelength. Each EPR source consists of a true zero order half wave plate (THWP) sandwiched by two identical 1mm-thick
beta barium borate (BBO) crystals with the type-II beam-like cutting type. Each BBO crystal produces the down-converted photons with the state of $|H_1V_2\rangle$, while the photon pair in the first BBO is transformed into $|V_1H_2\rangle$ by the THWP. Superposition of the SPDC processes in two BBOs is made by applying a temporal and spatial compensation crystal in each photon path, which makes the two possible ways of down conversion indistinguishable, leading to the final state $\left(|H_1V_2\rangle+|V_1H_2\rangle\right)/\sqrt{2}$,
where the horizontal (vertical) polarization denotes the logic qubit 0 (1). A HWP convert the entangled state in to  $\left(|HH\rangle+|VV\rangle\right)/\sqrt{2}$.

\textit{Four-qubit state generation.--} The four-qubit GHZ state is prepared by interfering the indistinguishable photons from EPR pairs. As shown in Fig.~\ref{fig:4GHZ}, one of the photons in each EPR pair is directed into the PBS, which behaves as a parity check operator $|00\rangle\langle00|+|11\rangle\langle11|$ by postselecting the case where both of the photons are transmitted or reflected by PBS. With the overlap of arriving photons in PBS, the Hong-Ou-Mandel interference occurs. By entangling the two identical sources, The four-qubit GHZ state $|\mathrm{GHZ_4}\rangle=\left(|0\rangle^{\otimes 4}+|1\rangle^{\otimes 4}\right)/\sqrt{2}$ is prepared. 

The biseparable state that is used to demonstrate the false positives of standard Mermin witness is comprised of one qubit state and a three-qubit GHZ state. The state preparation is based on the setup (optical elements on blue plate in Fig. \ref{fig:4GHZ}) where one PBS is applied to one of the sources to postselect the separable state $|00\rangle$ generated by only one of BBO. The following local unitaries are applied to transform the separable state into  $U_1\otimes U_2|01\rangle=\frac{1}{2}\left(|0\rangle+e^{i\pi/4}|1\rangle\right)\left(|0\rangle+e^{-i\pi/4}|1\rangle\right)$. Similarly as four-qubit GHZ, the separable state interferes with the other EPR source via Hong-Ou-Mandel effect on PBS, yielding the desired state $|\psi\rangle=(|0\rangle+\mathrm{e}^{i\pi/4}|1\rangle)(|0\rangle^{\otimes 3}+\mathrm{e}^{-i\pi/4}|1\rangle^{\otimes 3})/2$.

In order to achieve high fidelity, we use relatively low pump powers (17 mw) to suppress noise from higher-order emission events. This yields an approximate fourfold rate of 2.5 Hz. 

\section{Error analysis of lab observables}\label{app:error_analysis}

Characterizing the imperfection of the implemented measurement (lab observables) is a prerequisite for developing the corrected bi-separable bound of our protocol. In this section, the theoretical model of measurement imperfection is given.
\begin{table*}
\begin{ruledtabular}
\begin{tabular}{l|cccccc|c}
State preparation& $|H\rangle$& $|V\rangle$& $|D\rangle$& $|A\rangle$& $|R\rangle$& $|L\rangle$& Fidelity\\
\colrule
Projector $|D\rangle$ & 183295 & 156835 & 346731& 96& 164614& 182170& 0.9994$\pm$0.00001\\
Projector $|A\rangle$ & 166765 & 191123 & 180& 348779& 182989& 164313& 0.9994$\pm$0.00001\\
Projector $|R\rangle$ & 176340 & 170533 & 190799& 153178& 342721& 1586& 0.9976$\pm$0.00005\\
Projector $|L\rangle$ & 168310 & 175173 & 154604& 191831& 1690& 338662& 0.9977$\pm$0.00006\\
Projector $|H\rangle$ & 362365 & 188 & 182419& 181078& 186949& 179468& 0.9997$\pm$0.00002\\
Projector $|V\rangle$ & 144 & 370981 & 180420& 179439& 178401& 185962& 0.9998$\pm$0.00002\\
\end{tabular}
\end{ruledtabular}
\caption{\label{tab:measurement imperfection}%
\textbf{Raw data of measurement tomography}. }
\end{table*}

\begin{figure*}
\begin{center}
    \includegraphics[width=1\textwidth]{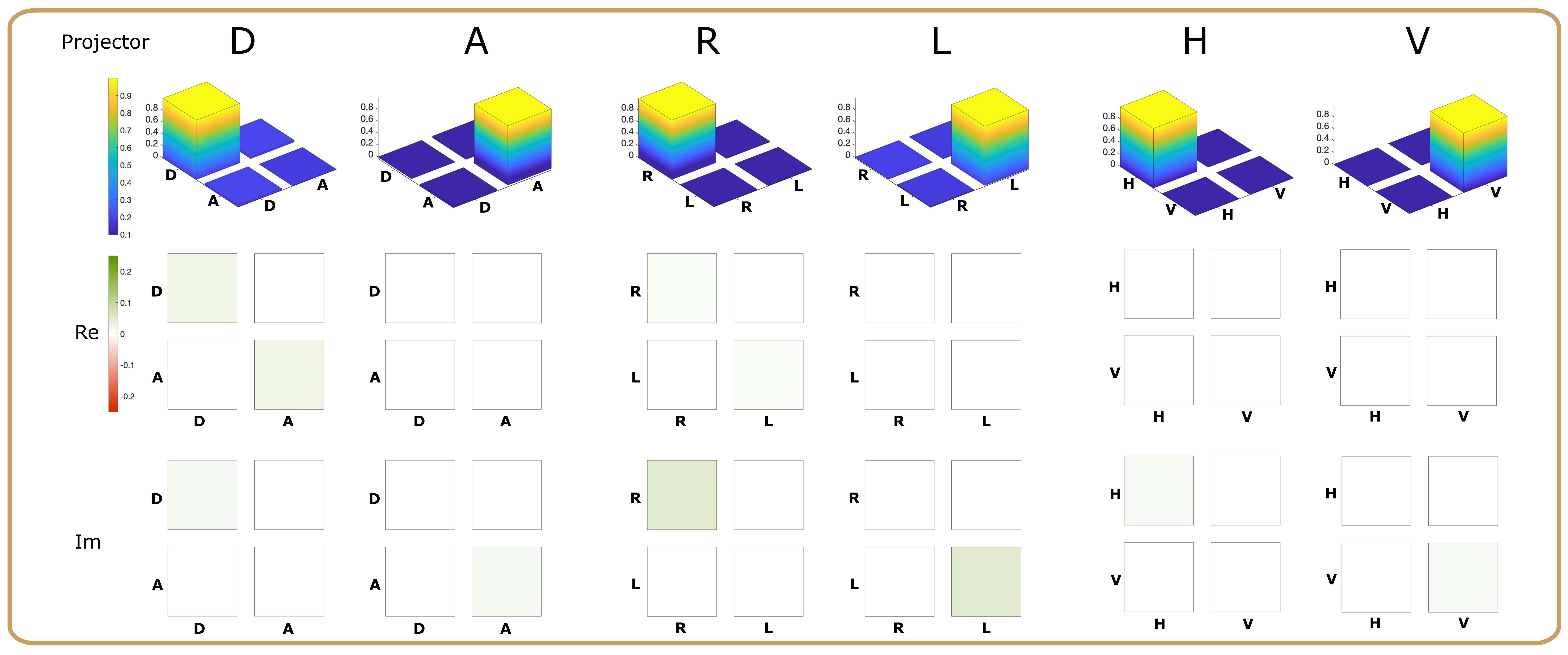}
    \caption{\textbf{Tomographic results of  projectors}. Here the projector $(D, A)$ basis corresponds to the eigenbasis of observable $X$ with the eigenvalue of $(+1, -1)$, while the $(R, L)$ corresponds to observable $Y$ and the $(H, V)$ corresponds to observable $Z$. The real part of the tomographic matrix is depicted in the first raw. For a clear comparison with ideal projectors, the difference between the tomographic projectors and ideal projectors are presented in the second (real part) and third (imaginary part) raw. }
    \label{fig:meatomo}
\end{center}    
\end{figure*}
The systematic error propagation takes the finite extinction ratio of PBS, the misalignment of the motorized rotation stage for waveplates, and the imperfect retarder of waveplates into account. An ideal PBS would transmit horizontal polarization (encoded as $|0\rangle$) and totally reflect vertical polarization (encoded as $|1\rangle$). Then the target measurement basis, defined as a combination of QWP, HWP and transmitted outcome of PBS, would be $|\psi\left(\theta,\phi\right)\rangle$ ($|\psi^\bot\left(\theta,\phi\right)\rangle$ for reflected outcome), where the $\theta,\phi$ represent the parameters in Bloch space that define the direction of the target basis, i.e. $|\psi\left(\theta,\phi\right)\rangle=
\begin{bmatrix}
    \mathrm{cos}\left(\theta\right)\\
    \mathrm{sin}\left(\theta\right)e^{i\phi}
\end{bmatrix}$.
However, due to the finite extinction ratio of the PBS, there is a small amount of polarization directed by PBS the opposite way, thus a realistic measurement of the transmitted port of PBS that actually performed is described as a POVM operator $\tilde{P}$

\begin{equation}
    \begin{aligned}
        \tilde{P} & = \sum_{i=0,1}\tilde{M_i} \tilde{M_i}^\dag \\
        \tilde{M_0} & = \gamma|\tilde{\psi}\left(\tilde{\theta,}\tilde{\phi}\right)\rangle \langle\tilde{\psi}\left(\tilde{\theta,}\hat{\phi}\right)|\\
        &=\gamma Q^{\dag}(\tilde{\alpha},\tilde{\eta}_q)H^{\dag}(\tilde{\beta},\tilde{\eta}_h)|0\rangle \langle 0|H(\tilde{\beta},\tilde{\eta}_h)Q(\tilde{\alpha},\tilde{\eta}_q)\\
        \tilde{M_1} & = \left(1-\gamma\right)|\tilde{\psi}^{\bot}\left(\tilde{\theta,}\tilde{\phi}\right)\rangle \langle\tilde{\psi}^{\bot}\left(\tilde{\theta,}\hat{\phi}\right)|\\
        & =\left(1-\gamma\right)Q^{\dag}(\tilde{\alpha},\tilde{\eta}_q)H^{\dag}(\tilde{\beta},\tilde{\eta}_h)|1\rangle \langle 1|H(\tilde{\beta},\tilde{\eta}_h)Q(\tilde{\alpha},\tilde{\eta}_q)
    \end{aligned}
    \label{eq:measurement operator}
\end{equation}
Here the tilde is used to denote the basis (with angles) actually implemented in the laboratory as opposed to the idealized target basis (with corresponding angles). The $Q(\tilde{\alpha},\tilde{\eta}_q)$ ($H(\tilde{\beta},\tilde{\eta}_h)$) describe the operations of QWP (HWP) when setting into an angle of $\tilde{\alpha}$ ($\tilde{\beta}$) while the actual retarder of it is $\tilde{\eta}_q$ ($\tilde{\eta}_h$), and $\gamma$ is the probability of correct reaction of PBS determined by the extinction ratio. It is obvious that the direction of practical basis in Bloch sphere $(\hat{\theta},\hat{\phi})$ is determined by the implemented angles of the waveplates $(\tilde{\alpha},\tilde{\beta})$, as well as the retarders of waveplates $(\tilde{\eta}_q,\tilde{\eta}_h)$. The imprecision of the quantities is modeled as
\begin{equation}
    \begin{aligned}
     \tilde{\theta}&=\theta+\Delta \theta \\
     \tilde{\phi}&=\phi+\Delta \phi \\
     \tilde{\alpha}&=\alpha+\Delta \alpha \\
     \tilde{\beta}&=\beta+\Delta \beta. \\
     \tilde{\eta}_q&=\eta_q+\Delta \eta_q \\
     \tilde{\eta}_h&=\eta_h+\Delta \eta_h 
    \end{aligned}
    \label{eq:angles}
\end{equation}
Here the desired retarders of QWP and HWP are $\eta_q=\frac{\pi}{2}, \eta_h=\pi$. The imprecision in the measuring direction in Bloch space can be estimated through the error propagation function of systematic error of optical elements
\begin{equation}
    \begin{aligned}
        \Delta \theta=\left|\frac{\partial \theta}{\partial \alpha}\right|\Delta\alpha+ \left|\frac{\partial \theta}{\partial \beta}\right|\Delta \beta+ \left|\frac{\partial \theta}{\partial \eta_q}\right|\Delta \eta_q+ \left|\frac{\partial \theta}{\partial \eta_h}\right|\Delta \eta_h\\
        \Delta \phi=\left|\frac{\partial \phi}{\partial \alpha}\right|\Delta\alpha+ \left|\frac{\partial \phi}{\partial \beta}\right|\Delta \beta+
        \left|\frac{\partial \phi}{\partial \eta_q}\right|\Delta \eta_q+ \left|\frac{\partial \phi}{\partial \eta_h}\right|\Delta \eta_h.
    \end{aligned}
    \label{eq:error propagation}
\end{equation}
In our case, the maximal misalignment of the setting angles by motorized rotation stages are $\Delta\alpha, \Delta\beta=0.4\pi/180$ according to the specification of motorized rotation stage, deviation of retarder $\Delta\eta_q=\Delta\eta_h=\pi/100$, and the high extinction ratio (>1000:1) PBS for both output port is chosen, leading to a $\gamma=0.999$ approximately. By substituting these parameters into Eq.~(\ref{eq:measurement operator})-(\ref{eq:error propagation}), one obtains the fidelity of the measurement operator via $F_{\tilde{P}}=\mathrm{Tr}\left(\tilde{P} \left|\psi\left(\theta,\phi\right)\rangle \langle\psi\left(\theta,\phi\right)\right|\right) =\gamma\left|\langle\tilde{\psi}(\tilde{\theta,}\hat{\phi})|\psi\left(\theta,\phi\right)\rangle\right|^2 + (1-\gamma)\left|\langle\tilde{\psi}^\bot(\tilde{\theta,}\hat{\phi})|\psi\left(\theta,\phi\right)\rangle\right|^2$. The theoretical model give rise to the fidelities of observables $\mathcal{F}_Z=99.89\%, \mathcal{F}_X=99.82\%, \mathcal{F}_Y=99.78\%$ which are quite close to our experimental results.


\section{Examples of four-partite device-independent entanglement witness}

Here we present the explicit examples of device-independent entanglement witness (DIEW) $I_{nm}$ in terms of four-qubit entangled state $n=4$, including the two settings $m=2$ and three settings $m=3$ per party \cite{barreiro2013demonstration}.

\begin{table*}
\begin{ruledtabular}
\begin{tabular}{cccc|c|cccc|c}
\multicolumn{5}{c|}{$s\equiv0(mod\,2)$} & \multicolumn{5}{c}{$s\equiv1(mod\,2)$}\\ \hline
$s_1$ & $s_2$ & $s_3$ & $s_4$ & $(-1)^{s/2}$ & $s_1$ & $s_2$ & $s_3$ & $s_4$ & $(-1)^{(s-1)/2}$\\ \hline
0 & 0 & 0 & 0 & 1 & 0 & 0 & 0 & 1 & 1\\ 
0 & 0 & 1 & 1 & -1 & 0 & 0 & 1 & 0 & 1\\ 
0 & 1 & 0 & 1 & -1 & 0 & 1 & 0 & 0 & 1\\ 
0 & 1 & 1 & 0 & -1 & 0 & 1 & 1 & 1 & -1\\ 
1 & 0 & 0 & 1 & -1 & 1 & 0 & 0 & 0 & 1\\ 
1 & 0 & 1 & 0 & -1 & 1 & 0 & 1 & 1 & -1\\ 
1 & 1 & 0 & 0 & -1 & 1 & 1 & 0 & 1 & -1\\ 
1 & 1 & 1 & 1 & 1 & 1 & 1 & 1 & 0 & -1\\ 
\end{tabular}
\end{ruledtabular}
\caption{\label{tab:I42}%
All possible measurement settings of $I_{42}$. }
\end{table*}

\begin{table*}
\begin{ruledtabular}
\begin{tabular}{cccc|c|cccc|c}
\multicolumn{5}{c|}{$s\equiv0(mod\,3)$} & \multicolumn{5}{c}{$s\equiv1(mod\,3)$}\\ \hline
$s_1$ & $s_2$ & $s_3$ & $s_4$ & $(-1)^{s/3}$ & $s_1$ & $s_2$ & $s_3$ & $s_4$ & $(-1)^{(s-1)/3}$\\ \hline
0 & 0 & 0 & 0 & 1 & 0 & 0 & 0 & 1 & 1\\ 
0 & 0 & 1 & 2 & -1 & 0 & 0 & 1 & 0 & 1\\ 
0 & 0 & 2 & 1 & -1 & 0 & 0 & 2 & 2 & -1\\ 
0 & 1 & 0 & 2 & -1 & 0 & 1 & 0 & 0 & 1\\ 
0 & 1 & 1 & 1 & -1 & 0 & 1 & 1 & 2 & -1\\ 
0 & 1 & 2 & 0 & -1 & 0 & 1 & 2 & 1 & -1\\ 
0 & 2 & 0 & 1 & -1 & 0 & 2 & 0 & 2 & -1\\ 
0 & 2 & 1 & 0 & -1 & 0 & 2 & 1 & 1 & -1\\ 
0 & 2 & 2 & 2 & 1 & 0 & 2 & 2 & 0 & -1\\
1 & 0 & 0 & 2 & -1 & 1 & 0 & 0 & 0 & 1\\ 
1 & 0 & 1 & 1 & -1 & 1 & 0 & 1 & 2 & -1\\ 
1 & 0 & 2 & 0 & -1 & 1 & 0 & 2 & 1 & -1\\ 
1 & 1 & 0 & 1 & -1 & 1 & 1 & 0 & 2 & -1\\ 
1 & 1 & 1 & 0 & -1 & 1 & 1 & 1 & 1 & -1\\ 
1 & 1 & 2 & 2 & 1 & 1 & 1 & 2 & 0 & -1\\ 
1 & 2 & 0 & 0 & -1 & 1 & 2 & 0 & 1 & -1\\
1 & 2 & 1 & 2 & 1 & 1 & 2 & 1 & 0 & -1\\ 
1 & 2 & 2 & 1 & 1 & 1 & 2 & 2 & 2 & 1\\ 
2 & 0 & 0 & 1 & -1 & 2 & 0 & 0 & 2 & -1\\ 
2 & 0 & 1 & 0 & -1 & 2 & 0 & 1 & 1 & -1\\ 
2 & 0 & 2 & 2 & 1 & 2 & 0 & 2 & 0 & -1\\ 
2 & 1 & 0 & 0 & -1 & 2 & 1 & 0 & 1 & -1\\ 
2 & 1 & 1 & 2 & 1 & 2 & 1 & 1 & 0 & -1\\ 
2 & 1 & 2 & 1 & 1 & 2 & 1 & 2 & 2 & 1\\ 
2 & 2 & 0 & 2 & 1 & 2 & 2 & 0 & 0 & -1\\
2 & 2 & 1 & 1 & 1 & 2 & 2 & 1 & 2 & 1\\ 
2 & 2 & 2 & 0 & 1 & 2 & 2 & 2 & 1 & 1\\ 
\end{tabular}
\end{ruledtabular}
\caption{\label{tab:I43}%
All possible measurement settings of $I_{43}$. }
\end{table*}

Recalling the general expression
\begin{equation}
    I_{nm}=\sum\limits_{s\equiv0(mod\,m)}(-1)^{s/m}E_s+\sum\limits_{s\equiv1(mod\,m)}(-1)^{(s-1)/m}E_s
\end{equation}
where $E_s=\sum_{\vec{\mathbf{r}}}(-1)^r P(\vec{\mathbf{r}}|\vec{\mathbf{s}})$ is the $n-$partite correlator with the vector of input $\vec{\mathbf{s}}=(s_1, s_2, \cdots s_n)\in \{0,1,\cdots m-1\}$ and output $\vec{\mathbf{r}}=(r_1, r_2, \cdots r_n)\in \{0,1\}$. The $s=\sum_j s_j$ and $r=\sum_j r_j$ are the sum over all the indices. The corresponding conditional probability satisfies $\sum_{\vec{\mathbf{r}}}P(\vec{\mathbf{r}}|\vec{\mathbf{s}})=1$. Any quantum correlation comes from the bipartition of the state are bounded by the biseparable bound $I_{nm}\leq 2m^{n-2}cot(\pi/2m)\equiv B_{nm}$. In principle, the DIEW $I_{nm}$ makes no explicit restrict on the local lab observable that applied to each party. Nonetheless, the practically observed correlations is based on some projecting operators $O_{r_i|s_i}^i$ acting on $i-$th qubit with the corresponding setting $s_i$ and outcome $r_i$. The probability distribution of the projecting operators is described as $P(\vec{\mathbf{r}}|\vec{\mathbf{s}})=\mathrm{Tr}\left[\bigotimes_{i=1}^n O_{r_i|s_i}^i \rho\right]$, where the $\rho$ is the $n-$partite quantum state. To be concrete, we specify the optimal observables for the genuine entangled n-qubit state 
\begin{equation}
    \begin{aligned}
        cos(\theta_{s_i})\sigma_x^i+sin(\theta_{s_i})\sigma_y^i\\
        \theta_{s_i}=-\frac{\pi}{2mn}+s_i\frac{\pi}{m}.
    \end{aligned}
    \label{eq:DIEWobservable}
\end{equation}
that maximize the evaluation of DIEW $\mathrm{max} \ I_{nm}=2m^{n-1}cos(\pi/2m)$.

For $n=4$ qubit with each qubit having the $m=2$ measurement settings, the combination of the input choices would be $2\times2\times2\times2=16$. All these possible measurements contributes to the evaluation of $I_{42}$. Depending on the value of $s$ modulo 2, we present the considered 16 measurement settings in Table \ref{tab:I42}, the bi-separable bound that associates with the $I_{42}$ is $B_{42}=8$.

For $n=4$ qubit with each qubit having the $m=3$ measurement settings, in total the possible input choices would be $3\times3\times3\times3=81$. Among these measurement setting, only 54 out of 81 contributes to the evaluation of $I_{43}$. We present the valid 54 measurement settings in Table \ref{tab:I43}, with the bi-separable bound $B_{43}=18\sqrt{3}$ associated to the $I_{43}$.


\section{Noise tolerance and robustness analysis}\label{app:robustness}

The robustness of entanglement witnesses against various types of noise is an important property and relevant for experimental use. 
We compare the Mermin witness of Eq. (4)
and the stabilizer witness of Eq. (7)
with measurement imprecisions under two types of noise, namely white noise or depolarizing noise and dephasing noise on the state that is distributed.
For the white noise we assume that the state that is actually measured is described by
\begin{align}\label{eq:noise_white}
    \rho_{\text{iso}}(p)=p\ketbra{GHZ}+\frac{1-p}{2^n}\mathds{1}.
\end{align}
In the case of dephasing noise, the state becomes
\begin{align}\label{eq:noise_dephasing}
    \rho_{\text{deph}}(p)=&p\ketbra{GHZ}
    +(1-p)\ketbra{GHZ^-},
\end{align}
with $|GHZ^-\rangle=(|0\dots0\rangle-|1\dots1\rangle)\sqrt{2}$.
To calculate the threshold level of noise where entanglement detection becomes impossible with a given inaccuracy $\varepsilon$ we have to distinguish two cases.
First, we assume the best case scenario, where the measurements are precise. This corresponds to a situation where, although we expect a given imprecision in the measurements and thus correct the biseparable bound accordingly, the actual measurements are executed perfectly.
Second, we assume the worst case, namely that the measurements are actually as imprecise as allowed by the fidelity bound.

\begin{figure}[t]
     \includegraphics[width=1\columnwidth]{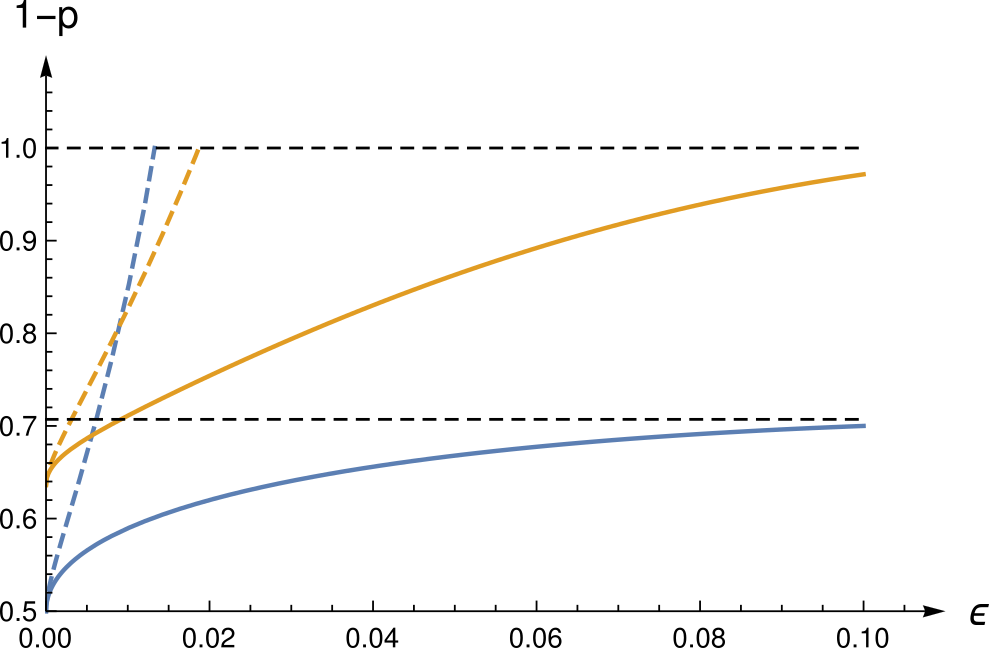}\\
     (a)\\
     \includegraphics[width=1\columnwidth]{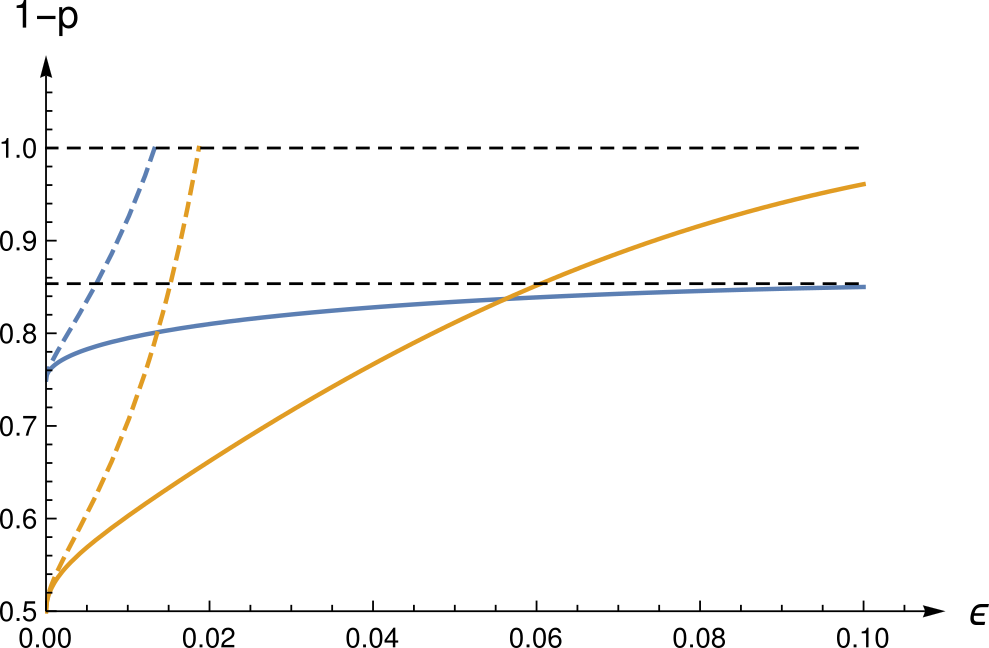}\\
     (b)\\
        \caption{
        \textbf{Comparison of the robustness under noise.}
        The plot shows $p$, where $1-p$ is the maximally tolerated noise level, versus the bound $\varepsilon$ on the misalignment of the measurements for the Mermin witness (blue) and the stabilizer witness (yellow) under white noise (above) and dephasing noise (below).
        The solid line shows the minimal value of $p$ that is needed to detect entanglement under the assumption of imprecise measurements, but where actually perfect measurements are performed.
        The dashed line shows the minimal value of $p$ that is needed to detect entanglement under the same assumption, with the worst possible measurements constraint to this assumption.
        These two lines mark two extreme cases, realistically the visibility lies somewhere between them.
        }
    \label{fig:rob4}
\end{figure}

For both cases, we can calculate the expectation value of the witnesses and compare it with the biseparable bounds.
Assuming precise measurements, the expectation value of both witness-operators scales linearly with the level of depolarizing noise, $\tr(\mathcal{M}^{(4)}\rho_{\text{iso}}(p))=8p$ and $\tr(\mathcal{W}^{(4)}\rho_{\text{iso}}(p))=11p$.
This means that the thresholds where entanglement detection becomes impossible are just the renormalised biseparable bounds $p=\mathbf{M}^{(4)}_{\text{bisep}}(\varepsilon)/8$ and $p=\mathbf{W}^{(4)}_{\text{bisep}}(\varepsilon)/11$, shown as a solid line in blue for the Mermin witness and in yellow for the stabilizer witness in Fig.~\ref{fig:rob4}(a).
Here the biseparable bound increases to account for imprecise measurements, but the expectation value of the witness remains the same since the actual measurements are performed perfectly.
In case that the measurements can be imprecise, we find the lower bound $\tr(\mathcal{M}_\varepsilon^{(4)}\rho_{\text{iso}}(p))\ge8p(1-8q^2+8q^4)$ and $\tr(\mathcal{W}_\varepsilon^{(4)}\rho_{\text{iso}}(p))\ge p(3-24q^2+32q^4)$ and with this $p=\mathbf{M}^{(4)}_{\text{bisep}}(\varepsilon)/(8(1-8q^2+8q^4))$ and $p=\mathbf{W}^{(4)}_{\text{bisep}}(\varepsilon)/(3-24q^2+32q^4)$, shown as dashed lines in blue and yellow in Fig.~\ref{fig:rob4}(b) respectively.
Here we observe two effects: the separable bound increases to account for the imprecision in the measurements and the expectation value of the witness decreases due to this imprecision.

The analog calculation can be done for the dephasing noise model. For the best case scenario we compute $p=(\mathbf{M}^{(4)}_{\text{bisep}}(\varepsilon)-8)/16$ and $p=(\mathbf{W}^{(4)}_{\text{bisep}}(\varepsilon)-3)/8$ for the perfect measurements, shown as a solid line in blue for the Mermin witness and in orange for the stabilizer witness in Fig.~\ref{fig:rob4}(b).
For the worst case we have $p=\mathbf{M}^{(4)}_{\text{bisep}}(\varepsilon)/(16(1-8q^2+8q^4))+1/2$ and $p=(\mathbf{W}^{(4)}_{\text{bisep}}(\varepsilon)+3(1-12q^2+10q^4))/(2(3-30q^2+31q^4))$ for the most imprecise measurements, shown as a dashed lines in Fig.~\ref{fig:rob4}(b).

We see that, depending on the type of noise we encounter, one witness is more resistant than the other. This behaviour changes with increasing imprecision in the performed measurements.
Noteworthy is that the Mermin witness loses less noise tolerance with increasing imprecision, when the potential imprecision of the measurements is included in the analysis but actual perfect measurements are performed. That is, while for both witnesses the visibility decreases, this rate is lower for the Mermin witness. In the case of dephasing noise, this is more evident: initially, the Mermin witness has a lower visibility, but with increasing imprecision, this changes and the visibility becomes higher than for the stabilizer witness. This makes sense since for the Mermin witness there is a gap between the quantum bound and the bilocal bound. Therefore the witness is more robust against misalignments in the measurement direction, even assuming a high inaccuracy does still allow for potential entanglement detection.
In the case of actually imprecise measurements, the visibility drops sharply for both witnesses and types of noise, here we see no clear advantage for either witness.


\end{document}